\documentclass{article}

\usepackage[english]{babel}

\usepackage[letterpaper,top=2cm,bottom=2cm,left=3cm,right=3cm,marginparwidth=1.75cm]{geometry}

\usepackage{amsmath}
\usepackage{amssymb}
\usepackage{amsfonts}
\usepackage{graphicx}
\usepackage[colorlinks=true, allcolors=blue]{hyperref}
\usepackage{mathrsfs}
\usepackage{textcomp}
\RequirePackage{amsthm}
\RequirePackage{latexsym}
\usepackage{array}
\usepackage[usenames, dvipsnames]{color}
\usepackage[font=small]{caption}
\usepackage{cite}
\usepackage[utf8]{inputenc}
\usepackage{multicol}
\usepackage{bbold}

\usepackage{pythonhighlight} 

\newtheorem*{definition*}{Definition}
\newtheorem{theorem}{\bf Theorem}

\newtheorem{Theorem}{Theorem}[section]

 { \theoremstyle{definition}
\newtheorem{Definition}[Theorem]{Definition}

 }

\newcommand{\cL}{\mathcal{L}}

\newcommand{\gpred}{g_{\text{pr}}}

\newcommand{\vCY}{\text{Vol}_{\text{CY}}}

\renewcommand{\Re}{{\rm Re}\;} 
\renewcommand{\Im}{{\rm Im }\;}

\title{
\bf Lectures on Numerical and Machine \\Learning Methods for Approximating \\Ricci-flat Calabi-Yau Metrics}
\author{Lara B. Anderson, James Gray, Magdalena Larfors}

\begin{document}
\maketitle

\begin{abstract}
Calabi--Yau (CY) manifolds play a ubiquitous role in string theory. As a supersymmetry-preserving choice for the 6 extra compact dimensions of superstring compactifications, these spaces provide an arena in which to explore the rich interplay between physics and geometry. These lectures will focus on compact CY manifolds and the long standing problem of determining their Ricci flat metrics. Despite powerful existence theorems, no analytic expressions for these metrics are known.

In this lecture series we review numerical approximation methods for Ricci flat CY metrics. Our first aim is to give a brief overview of the mathematical framework underlying CY geometry, and the various metrics that CY manifolds admit. We will then discuss the three types of numerical methods that have been developed to compute Ricci-flat CY metrics: Donaldson's algorithm \cite{donaldson2005numerical}, functional minimization methods \cite{Headrick:2009jz}, and machine learning methods \cite{Ashmore:2019wzb,Anderson:2020hux,Jejjala:2020wcc,Douglas:2020hpv,Douglas:2021zdn,Larfors:2021pbb,Larfors:2022nep,Gerdes:2022nzr}. Due to the limited time/space we have, this will not be a comprehensive review, but instead we hope to give a brief survey and illustrate the essential tools, key ideas, and implementations of this rapidly advancing field.
\end{abstract}

\section{Lecture 1: Calabi-Yau Geometry}\label{lect1}
\subsection{Introduction}
The goal of today's lecture is to set the stage for the central topic of these lectures: using Machine Learning to solve (or more accurately, approximate the solution to) Einstein's equations for Ricci flat metrics on spaces with non-trivial topology. In particular, we are interested in geometries known as \emph{Calabi-Yau} manifolds and their metrics, which play a special role in string theory. In particular, this set of lectures is about solving partial differential equations in curved spaces. For this goal we clearly need some notions of geometry and also a little motivation as to \emph{why} this problem is interesting. So, to that end, let's begin by asking: ``What is a Calabi-Yau manifold and what is special about its metric?" 

A Calabi-Yau manifold is a compact K\"ahler manifold with a Ricci-flat metric (and thanks to Yau's theorem, this is equivalent to a K\"ahler manifold with vanishing first Chern class). The more formal definition is given in Section \ref{sec:cy} below. At the moment though, these terms may not mean much to you and as a result, we must provide some background for our discussion. There are a great many well-written and useful introductions to Calabi-Yau geometry (see e.g. \cite{Candelas:1987is,Green:1987mn,greene1997string,Anderson:2018pui}) and these lectures will not attempt to replicate those efforts. Instead, our goal is to provide as quick and concise a survey as possible of the key features of these geometries, with an eye to what will matter most to us in a machine learning/computational approach.

Within the subject of string theory, determining a Ricci-flat metric on a Calabi-Yau manifold is a very big prize indeed. To understand why, we need to begin by noting that string theory is a theory of fundamental interactions that is formulated in more than the $3+1$ dimensions of space and time that we observe in our universe. In fact, a lot more. In heterotic string theory for example, the underlying theory is defined in $10$-dimensions ($9$ of space and $1$ time-like dimension). The low energy effective theory can be determined by a Lagrangian of the form (see e.g. \cite{Candelas:1985en})
\begin{equation}\label{10Dlagrangian}
S_{10D} \sim \int \sqrt{-g}d^{10}x \left(R + \alpha F_{\mu \nu}F^{\mu \nu} + \ldots \right)
\end{equation}
for gauge theory in higher dimensions, coupled to gravity. If you haven't encountered metrics before stay tuned for Section \ref{kahler_metric} below, but for now note that the metric $g$ determines the volume form and all inner products/contractions in the integral above.

To reduce this to a $4$-dimensional theory that could potentially be relevant to our world, we could look for solutions of the form $\mathbb{R}^{1,3} \times X_6$ where $X_6$ is some \emph{compact} manifold. With such a choice, one could imagine performing the multi-dimensional integral over $X_6$ and effectively ``integrating out" the extra dimensions to obtain a $4$-dimensional Lagrangian theory
\begin{equation}
S_{4D} \sim \int \sqrt{-g} d^4x \left(R+  \tilde{\alpha} F_{\mu \nu}F^{\mu \nu}+\ldots \right)
\end{equation}
The process of obtaining $S_{4D}$ is referred to as a ``string compactification". Which compact geometries $X_6$ are allowed is determined by the symmetries of \eqref{10Dlagrangian} and its equations of motion. One such class of solutions are given by Calabi-Yau manifolds which admit Ricci flat metrics (see \cite{Green:1987mn}). Clearly, the form of such a Lagrangian depends strongly on $X_6$ and its metric, $g_{\mu\nu}$. A great many quantities that specify the physical theory are a function of this metric, including (but not limited to)
\begin{itemize}
\item The cosmological background of the theory (i.e. cosmological constant, etc)
\item Masses of fields/particles
\item Couplings of fields
\item Gauge couplings
\end{itemize}
Unfortunately, the analytic form of the relevant Ricci flat metrics for compact spaces are not known\footnote{For non-compact CY manifolds, there are analytic expressions for Ricci-flat metrics. The conifold metric is one famous example \cite{Candelas:1989js}. }. For 30 years, string theorists have devised clever ways of working around our ignorance of the fundamental geometry and managed to extract many interesting properties of string compactifications. However, in order to fully understand the effective physics of a string compactificaton to $4$-dimensions (and decide what it is useful for!), the question of $X_6$ and its metric must be addressed. 

This is a difficult problem mathematically (as we will discuss later, a Fields Medal was awarded to Yau for proving that a Calabi-Yau metric exists, but the proof is not constructive). Since analytic tools to generate the metric are sparse, it is natural to turn to numeric approaches. Some progress has been made in this area over a number of years (see e.g. \cite{donaldson2005numerical,Headrick:2009jz,Douglas:2006hz,Douglas:2006rr,Braun:2007sn,Braun:2008jp}), but a substantial new tool has arisen with machine learning algorithms. This has enabled some very exciting recent progress \cite{Ashmore:2019wzb,Anderson:2020hux,Jejjala:2020wcc,Douglas:2020hpv,Douglas:2021zdn,Larfors:2021pbb,Larfors:2022nep,Gerdes:2022nzr} and this is what we hope to give you a feel for in these lectures. We'll turn now to a brief review of what type of ``geometry" must be specified about $X_6$ in the cases of interest (Calabi-Yau manifolds) and what is meant by a metric in this case.

\subsection{Manifolds in a nutshell}
Here we will provide a lightning review of a few tools from differential geometry related to manifolds and metrics. For a more detailed introduction to these topics, we suggest looking at \cite{huybrechts} and \cite{Candelas:1987is} for Calabi-Yau manifolds in particular. 

We'll begin with the basic definition of a manifold. Recall that, 
\begin{Definition}
A real manifold, $M$, consists of the following information
\begin{itemize}
\item A collection of open sets $\{U_\alpha\}$ such that $\cup_{\alpha} U_{\alpha}=M$.
\item For each $U_\alpha$, there exists a morphism $\phi_{\alpha}: U_{\alpha} \to \mathbb{R}^n$.
\item If $U_\alpha \cap U_{\beta} \neq 0$ then $\phi_{\alpha}|_{U_\alpha \cap U_{\beta}} \circ \phi_{\beta}^{-1}|_{U_\alpha \cap U_{\beta}}$ is a diffeomorphism.
\end{itemize}
The collection $\{U_{\alpha},\phi_{\alpha}\}$ is called an \emph{atlas}. 
\end{Definition}

Note that complex manifolds are defined in a very similar way, except that we require mappings ${\bf z}_{\alpha}: U_{\alpha} \to \mathbb{C}^n$ and ${\bf z}_{a} \circ \bf{z}_{b}^{-1}$ is a \emph{holomorphic} function (i.e. a function of $z$ and not its conjugate) on ${\bf z}_{b}(U_{\alpha} \cap U_{\beta}) \subset \mathbb{C}^n$ for all $a,b$. The mapping itself ${\bf z}_{a}=(z_{a}^1,\ldots, z_{a}^n)$ is referred to as ``local holomorphic coordinates on the complex manifold and on patch overlaps, $U_{a}\cap U_{b}$, the transition functions $z_{a}^i=f^i_{ab}(z_b)$ are holomorphic. An atlas with these properties is referred to as a \emph{complex structure} on $M$. It is clear that every $n$-dimensional complex manifold is a $2n$-dimensional real manifold, but the converse is not true. For example, the spheres $S^{2n}$ are only complex for $n=1$.

In the context of string theory, we are interested frequently in compact, complex manifolds. It turns out that this first property of compactness can present challenges in the complex setting. It is straightforward to build a compact, algebraic submanifold of $\mathbb{R}^n$. For example the algebraic constraint
\begin{equation}
x^2 +y^2=1 \subset \mathbb{R}^2
\end{equation}
is a circle and clearly defines a compact, 1-dimensional submanifold of $\mathbb{R}^2$. But in $\mathbb{C}^n$ this turns out to be harder. In fact, $\mathbb{C}^n$ has no compact, complex submanifolds. This is the case since any global holomorphic function on a compact, complex manifold is a constant. Applying this to the coordinate functions yields the issue.

So we are forced to choose a different complex ``ambient space" in which to build the compact, complex manifolds of interest. For this purpose, it is helpful to consider a ``compactified" version of $\mathbb{C}^n$, known as \emph{complex projective space} and denoted $\mathbb{C}\mathbb{P}^n$ (frequently in the physics literature this is simply written as $\mathbb{P}^n$ with complexity assumed). Complex projective space is defined as the set of complex lines through the origin of $\mathbb{C}^{n+1}$.

Note that a line through the origin is defined by a single point and points, $z,w$ define the same line if and only if there exists $\lambda \in \mathbb{C}^*=\mathbb{C} - \{0\}$ such that $(Z^0,Z^1,\ldots,Z^n)=\lambda(W^0,\ldots, W^n)$. We write this as
\begin{align}
&\mathbb{P}^n=\frac{\mathbb{C}^{n+1}-\{0\}}{\mathbb{C}^*} & \text{or} && \frac{S^{2n+1}}{U(1)}
\end{align}
For $\mathbb{P}^n$, the coordinates $(Z^0,\ldots,Z^n)$ are called homogeneous coordinates and identified under the equivalence relation
\begin{equation}\label{pn_coords}
(Z^0,Z^1,\ldots,Z^n) \sim \lambda(Z^0,\ldots, Z^n)~.
\end{equation}
A good atlas can be obtained by the coordinate patches defined by
\begin{equation}
U_i=\{ \left[ Z\right] : Z^i \neq 0 \}
\end{equation}
This is the set of lines through the origin that do not pass through $Z_i$. Natural local coordinates on each patch $U_i$ are then $w_i^k=\frac{Z^k}{Z^i}$. In the tutorial this afternoon you will play with these patches and prove to yourself that with the definitions above, the transition functions on overlaps are holomorphic functions. 

It is worth mentioning at this stage that the set up above for complex projective space can very readily be generalized. For example, complex \emph{weighted projective space} can be defined by introducing a scaling relation $(Z^0, \ldots, Z^n) \sim (\lambda_0 Z^0,\ldots, \lambda_n Z^n)$ instead of \eqref{pn_coords}. More generally there can be multiple $\mathbb{C}^*$ scaling relations of the coordinates and this leads to the beautiful framework of toric geometry. The interested reader can explore further here \cite{cox_little_schenck}.

\subsubsection{Hermitian and K\"ahler Metrics}\label{kahler_metric}
With these basic definitions of manifolds in hand, we turn now to the object that is central to this series of lectures: the metric. Very briefly the metric encodes information about lengths in (generally curved) spaces. For example, in 3-dimensions, instead of the line element expected from the Pythagorean theorem, dot products, etc ($ds^2=dx^2+dy^2+dz^2$), the notion of length in the space could be more complicated (i.e. $ds^2=g_{AB}dx^A dx^B$) as illustrated schematically in Figure \ref{fig:metrics} below. In the context of physics however, frequently metrics of interest are those on spacetime geometries, $g_{\mu \nu}$ where $\mu,\nu=0, \ldots n$ with signature $(-,+, \ldots, +)$. For the case of CY compactifications however as described above, we will focus on the metric on the internal, compact directions of space (e.g. $X_6$) only (all six directions are space-like). We will also focus this discussion on the complex setting developed above. 

\begin{figure}[ht]
\centering
\includegraphics[width=0.7\textwidth]{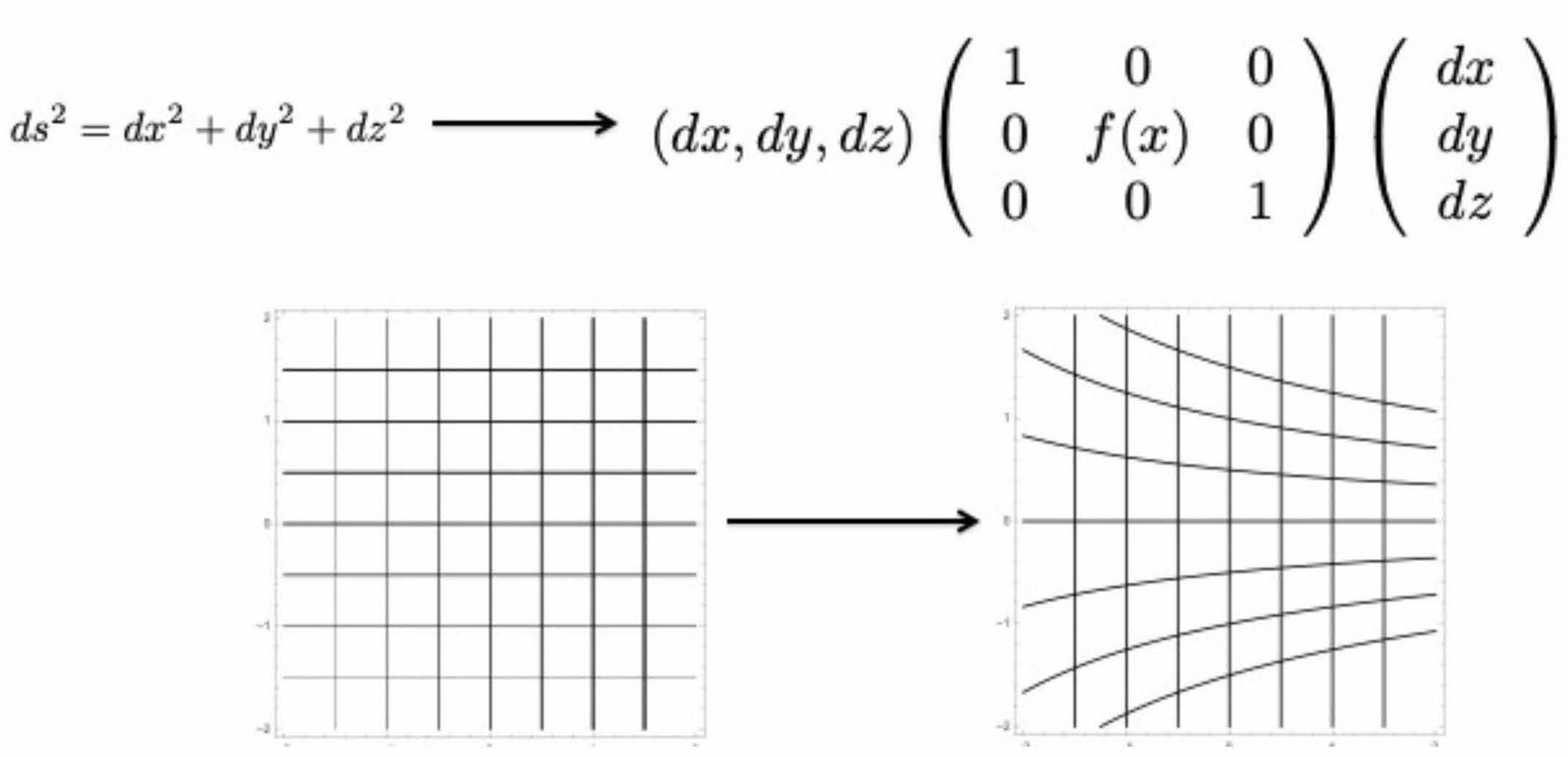}
\caption{\emph{An illustration of how a non-trivial metric could change the notion of length in a $2$-dimensional space.}\label{fig:metrics}}
\end{figure}

\begin{Definition}
A \emph{Hermitian metric} on a complex manifold, $X$, is a covariant tensor field $g_{i\bar{j}}dz^i \otimes d\bar{z}^{\bar{j}}$. In local coordinates, $g_{ij}=g_{\bar{i}\bar{j}}=0$, $g_{j\bar{i}}=\overline{g_{i\bar{j}}}$ and $g_{i\bar{j}}V^i \bar{V}^{\bar{j}} \geq 0$ for all $V^i$ (i.e. the inner product is positive semi-definite). It follows that this metric defines a (real) $(1,1)$ form
\begin{equation}\label{kahler_form}
J=i(g_{i\bar{j}}dz^i \wedge d\bar{z}^{\bar{j}})
\end{equation}
which is called the \emph{Hermitian form} or sometimes the fundamental form. Furthermore if this form satisfies the constraint that it is closed (i.e. that $dJ=0$) then it is called a \emph{K\"ahler form} and the manifold $X$ is called a K\"ahler manifold.
\end{Definition}

The condition $dJ=0$ can be expanded as
\begin{equation}\label{eq:hmetric}
dJ=i(\partial + \bar{\partial})g_{i\bar{j}}dz^i \wedge d\bar{z}^{\bar j}=0
\end{equation}
which implies that $\frac{\partial g_{i\bar{j}}}{\partial z^l}=\frac{\partial g_{l\bar{j}}}{\partial z^i}$, etc. This leads to the important fact that
\begin{equation}
g_{i\bar{j}}=\frac{\partial^2 K}{\partial z^i \partial \bar{z}^{\bar j}}~~~~,~~g_{ij}=g_{\bar{i}\bar{j}}=0
\end{equation}
That is, the K\"ahler metric is fully determined by the (locally defined) function $K=K(z,\bar{z})$ called the \emph{K\"ahler potential}.

It is worth noting that many notable/simple complex manifolds are also K\"ahler. In particular, complex projective space ($\mathbb{C}\mathbb{P}^n$ defined above) is K\"ahler. This fact can be important for approaches to numerically constructing metrics, as we will see later in this lecture series. For now, we will catalog several useful properties of K\"ahler manifolds (see e.g. \cite{Candelas:1987is} for more details).

The volume form of an $n$-dimensional K\"ahler manifold can be expressed in terms of the K\"ahler form as
\begin{equation}
\frac{J^n}{n!}\sim det(g)dz^1 \wedge \ldots dz^n\wedge d\bar{z}^1\wedge \ldots \wedge d\bar{z}^n
\end{equation}
and the Riemann tensor takes the simple form
\begin{equation}
R_{i\bar{j}k\bar{l}}=g_{k\bar{l},i\bar{j}}=-g^{p\bar{q}}g_{k\bar{q},i}g_{p\bar{l},\bar{j}}
\end{equation}
which in turn determines the Ricci tensor
\begin{equation}
R_{i\bar{j}}=-R_{i\bar{j}l\bar{k}}=-\partial_i\bar{\partial}_{\bar{j}}(\log(\det (g)))
\end{equation}
and the first Chern class of the manifold $X$
\begin{equation}
c_1(X)=\frac{i}{2\pi}R_{k\bar{l}}dz^k\wedge d\bar{z}^{\bar{l}}=\frac{1}{2\pi}R=-\frac{i}{2\pi}\partial\bar{\partial} \log (\det (g_{k\bar{l}}))~.
\end{equation}
With these definitions in place, we are ready to return to our discussion of Calabi-Yau manifolds and define things more carefully. 

\subsection{Calabi-Yau manifolds}\label{sec:cy}
With the definitions above, we can now understand better the definition of a Calabi-Yau manifold: a complex, K\"ahler manifold with $c_1(X)=0$.  In 1978, by studying complex Monge-Ampere equations (see Section \ref{monge_amp} below), Yau proved a conjecture of Calabi (posed in 1954). This and other important work led to Yau's Fields medal in 1982.

\begin{theorem} (\cite{Yau:1978cfy,Yau:1986gu}):
If $X$ is a complex, K\"ahler manifold with $c_1(X)=0$ and some K\"ahler form $J$, then there exists a unique, Ricci-Flat metric on $X$ whose K\"ahler form $J'$ is in the same cohomology class as $J$ (i.e. $J'=J+\partial \bar{\partial} \phi$).\label{yau}
\end{theorem}
This definition is useful, but as Yau also noted in the proof of his famous theorem, we will see that packaging the condition of a Ricci-flat metric in other terms can simplify the problem. Much of the structure of a Calabi-Yau manifold is encoded in two canonical forms. The first of these, the K\"ahler form, we have encountered already in \eqref{kahler_form}. The next is called the ``holomorphic top form". An $n$-dimensional Calabi-Yau manifold admits a nowhere vanishing, covariantly constant $(n,0)$ form, denoted by $\Omega$. In the case of a CY 3-fold (our primary interest in these lectures) the holomorphic 3-form can be constructed from the single covariantly constant spinor, $\epsilon$, which defines a manifold of $SU(3)$-holonomy \cite{BSMF_1955__83__279_0}
\begin{equation}
\Omega_{ijk}=\epsilon^T\gamma_{ijk}\epsilon
\end{equation}
with $\gamma_{ijk}$ the anti-symmetric product of $\gamma$-matrices satisfying
\begin{equation}
\{\gamma_i,\gamma_j\}=\{\gamma_{\bar{i}},\gamma_{\bar{j}}\}=0~~,~~\{\gamma_i,\gamma_{\bar{j}}\}=2g_{i\bar{j}}
\end{equation}
The fact that $\Omega$ is holomorphic (i.e. that $\bar{\partial}_{\bar{j}}\Omega_{i_{1},\ldots i_{n}}=0$) follows from $\Omega$ being covariantly constant.

The three-form $\Omega$ is essentially unique. Given $\Omega$, assume that there exists another 3-form $\Omega'$ with the same properties. Then since top forms must be proportional $\Omega'=f\Omega$ with a $f$ a non-singular function. But since $\bar{\partial} \Omega'=0$ by assumption, it follows that $f$ must be a holomorphic function. However, on a compact manifold this implies that $f$ is a constant. Later in this lecture we'll see an explicit form for the 3-form $\Omega$ for certain algebraic constructions of Calabi-Yau threefolds.

\subsubsection{K\"ahler geometry and the Monge-Ampere equations}\label{monge_amp}
The central goal of this lecture series is to use machine learning algorithms to attack this equation
\begin{equation}
R_{i\bar{j}}=-\partial_i\bar{\partial}_{\bar{j}}\log(\det(g))=0
\end{equation}
with $g_{i\bar{j}}=\partial_i \bar{\partial}_{\bar{j}} K$ defined by a K\"ahler potential $K$. Observe that this is a $4$-th order, very non-linear partial differential equation for $K$ in six (real) dimensions. As far as numerically solving PDEs goes, this is fairly awful. To make this tractable (both analytically and numerically) it is worth re-formulating this problem by following the approach used by Yau in his famous proof. Namely, to utilize the canonical differential forms $(J, \Omega)$ available on a Calabi-Yau manifold. Here is a sketch of the idea.

By the Calabi-Yau conjecture, start with \emph{any} K\"ahler metric $g$ on the CY 3-fold, $X$ and its associated K\"ahler form $J_g$. Then the Ricci-flat metric, $g_{CY}$ and its K\"ahler form $J$ can be written as
\begin{equation}\label{J_def}
J=J_g+ \partial \bar{\partial}\phi~.
\end{equation}
Observe then that there exists two ways of building a volume form on $X$:
\begin{equation}
\label{monge_ampere}
J\wedge J\wedge J= \kappa \Omega \wedge \bar{\Omega}~~~(\kappa \in \mathbb{C})
\end{equation}
where the LHS of the equation above was defined in \eqref{kahler_form} and the RHS follows from the properties of $\Omega$ outlined above. Using \eqref{J_def} we are lead to \eqref{monge_ampere}, a complex equation of \emph{Monge-Ampere type} for $\phi$ which is (only) second order. This equation was attacked using tools of differential geometry and analysis by Yau \cite{Yau:1978cfy}. As we will see in subsequent lectures, it is also precisely this equation that provides us a handy foothold into machine implementation.

\subsubsection{Calabi-Yau Moduli}\label{sec:moduli}
Calabi-Yau manifolds come equipped with rich and intricate moduli spaces. The word ``moduli space" can refer to fluctuations of the metric as we will see below, but also to vacuum spaces of fields in the physical theory and these two notions are deeply linked. For now we will focus on the geometric notion. See \cite{Green:1987mn,Hubsch:1992nu} for a deeper dive into these spaces.

If $X$ is a CY 3-fold then the Ricci-flatness condition guarantees that $R_{i\bar{j}}(g)=0$. We can consider perturbing the metric (i.e. $g \to g+\delta g$) such that the Ricci-flatness condition is still preserved
\begin{equation}
\delta g=\delta g_{ij}dz^i dz^j + \delta g_{i\bar{j}}dz^id\bar{z}^{\bar{j}}+c.c.
\end{equation}
Up to complex conjugation these fluctuations come in two types ($\delta g_{ji}$ and $\delta g_{i\bar{j}}$). The latter at least manifestly preserves the form of the initial K\"ahler metric. Plugging this fluctuation in to the Ricci-flatness condition leads to the constraint that $\delta g_{i\bar{j}}$ is harmonic. That is
\begin{equation}
\delta g_{i\bar{j}} \in H^{1,1}(X)
\end{equation}

Such deformations of the metric are known as \emph{Kahler moduli} and are responsible for varying volumes (including the overall volume) within the Calabi-Yau threefold.

The second type of metric deformation, $\delta g_{ij}$, seems to make the metric non-Hermitian (as the index structure is different than that required in \eqref{kahler_form}). However, there exist coordinate redefinitions (see e.g. \cite{Green:1987mn}) which can change this back to Hermitian from (i.e. only $i\bar{j}$-type indices) but they are \underline{not} holomorphic coordinate changes. As a result, the metric is Hermitian, but with respect to a different complex structure than we began with. Thus, deformations of this type are referred to as \emph{complex structure moduli}. It can be shown that
\begin{equation}
\Omega_{ijk}g^{k\bar{k}}\delta_{\bar{k}\bar{l}}dz^i\wedge dz^j \wedge d\bar{z}^{\bar{l}} \in H^{2,1}(X)
\end{equation}
We will see in the next Subsection that for simple algebraic constructions of Calabi-Yau manifolds (i.e. algebraic complete intersections) that the complex structure deformations parameterized by $H^{2,1}(X)$ correspond to varying the coefficients in polynomial defining equations (i.e. changing the algebraic constraint that defines the manifold).

In summary, for a Calabi-Yau threefold the two independent Hodge numbers ($h^{1,1}(X),h^{2,1}(X)$) count the number of independent metric fluctuations, $\delta g$, that preserve the Calabi-Yau condition.

\subsection{Building Simple Calabi-Yau Manifolds}
In this last bit of geometric introduction, we will outline simple algebraic constructions of Calabi-Yau threefolds. The idea is to utilize algebraic sub-varieties of simple ambient spaces (e.g. complex projective space, $\mathbb{P}^n$ defined above) to construct compact K\"ahler manifolds with $c_1(X)=0$.

To begin, let us consider a single polynomial constraint in $\mathbb{P}^n$. We will denote this hypersurface by $p(Z)=0$ where $Z_i$ are the homogeneous coordinates defined in \eqref{pn_coords}. Note that in these coordinates, it is imperative that the polynomial $p(Z)$ be of homogeneous degree (every term in the polynomial must be of uniform scaling under the action in \eqref{pn_coords}), otherwise the hypersurface is not well defined. For example a polynomial constraint of the form $
(Z^0)^2+(Z^1)=0$ in $\mathbb{P}^2$ does not yield a well-defined sub-variety (hint: what happens to this equation under the equivalence relation?) While by contrast $(Z^0)^2+(Z^1)^2=0$ is well defined (since each term scales as $\lambda^2$) and yields a one-(complex) dimensional sub-space of $\mathbb{P}^2$ (in this case a double cover of $\mathbb{P}^1$ itself).

It is straightforward to show that holomorphic, homogeneous polynomial constraints of this type yield K\"ahler sub-varieties of the ambient space. But what about the condition that $c_1(X)=0$ as required by Theorem \ref{yau}? It can be shown (and you'll explore this in the tutorial this afternoon) that the Calabi-Yau condition can be enforced simply for manifolds of this type.

If we begin with a simple K\"ahler metric on $\mathbb{P}^n$ known as the Fubini-Study metric
\begin{equation}\label{fubini_study}
g_{i\bar{j}}=\partial_i \bar{\partial}_{\bar{j}}\log(|Z^0|^2+ \ldots +|Z^n|^2)
\end{equation}
then one can try to do the natural thing and restrict this to the sub-space, $X$, defined by $p(Z)=0$. Suppose that the polynomial is of homogeneous degree $m$ in $\mathbb{P}^n$. Then a straightforward (but lengthy) curvature calculation yields 
\begin{equation}
R_{FS}|_X=((n+1)-m)g_{FS}+\text{(total derivative terms)}
\end{equation}
for the Ricci scalar. From this form, it is clear that 
\begin{equation}\label{cy_cond}
c_1(X)=0 ~~\text{when}~~~ m=n+1~.
\end{equation} 
We denote the resulting Calabi-Yau n-fold, $X$, by the data of its ambient space and degree as $X=\mathbb{P}^n\left[m\right]$.

The prototypical example of this type is a degree $5$ hypersurface in $\mathbb{P}^5$ (known as ``the quintic" threefold, $X=\mathbb{P}^4\left[5\right]$). As an explicit example, consider this simple polynomial (known as the Fermat polynomial)
\begin{equation}
p=(Z^0)^5 + \ldots +(Z^4)^5=0
\end{equation}
This is far from the only choice however for such a degree $5$ hypersurface. Indeed we could choose different constant coefficients for every one of the possible $\binom{5+4}{4}=126$ degree $5$ terms in such a polynomial
\begin{equation}\label{poly2}
p=\alpha_0(Z^0)^5+\alpha_1(Z^0)^4(Z^1)+\ldots
\end{equation}
Changing the coefficients $\alpha_a$ corresponds to changing the complex structure (and hence deforming the metric) as in Section \ref{sec:moduli} described above (see e.g. \cite{greene1997string} for further details). However, not all $126$ degrees of freedom above correspond to physical moduli. Indeed, there are 24 degrees of freedom corresponding to a general $PGL(5,\mathbb{C})$ coordinate transformation as well as an overall scale (which drops out of equations such as \eqref{poly2}). The result leaves 126-25=101 actual complex structure moduli for the quintic. In addition there is a single K\"ahler modulus, inherited from the restriction of the K\"ahler form on $\mathbb{P}^n$ (associated to the metric in \eqref{fubini_study}. As described in Section \ref{sec:moduli} above, the Hodge numbers of this Calabi-Yau manifold are then $(h^{1,1}(X),h^{2,1}(X))=(1,101)$.

There are many other manifolds that can be similarly constructed. For example, we could consider multiple defining equations in a single projective space (i.e. construct a complete intersection hypersurface). Using the same notation for ambient space and polynomial degree defined above we can catalog these so-called ``configuration matrices" as
\begin{equation}
\left[\mathbb{P}^4 | 5\right]~,~\left[\mathbb{P}^5|3,3\right]~,~\left[\mathbb{P}^5|4,2\right]~,~\left[\mathbb{P}^6|2,2,3\right]~,~\left[\mathbb{P}^7|2,2,2,2\right]
\end{equation}
These $5$ manifolds (known as `cyclic' Calabi-Yau threefolds) are the only CY 3-folds defined as algebraic subvarieties of a single projective space. However, we could readily generalize our  ambient space beyond a single projective space to, for example, products of projective spaces allowing us to define manifolds such as the following in $\mathbb{P}^1 \times \mathbb{P}^3$:
\begin{eqnarray}
&&\left[ \begin{array}{c|c}
\mathbb{P}_X^1 &2\\
\mathbb{P}_Z^3 &4
\end{array} \right]\\
p(z)&=&(X^0)^2(Z^0)^4+(X^0)(X^1)(Z^0)^2(Z^1)(Z^2) +\dots =0
\end{eqnarray}
More generally, to build a CY 3-fold within a product of $p$ projective spaces, $\mathbb{P}^{n_1} \times \ldots \mathbb{P}^{n_p}$, we need
\begin{equation}\label{ci}
\sum_{r=1}^p n_r - L = 3 \ .
\end{equation}
where $L$ is the number of defining equations, $p_j(z)=0$, forming the complete intersection. Each of the homogeneous polynomials $p_j$ which define the manifold can be characterized by its multi-degree ${\bf m}_j=(m_j^1,\ldots , m_j^m)$, where $m_j^r$ specifies the degree of $p_j$ in the homogeneous coordinates ${\bf z}^{(r)}$ of the factor $\mathbb{P}^{n_r}$ in $\mathcal{A}$.  A standard way to encode this information is by a so-called {\it configuration matrix}
\begin{equation}\label{cy-config}
\left[\begin{array}{c|cccc}
\mathbb{P}^{n_1} & m_{1}^{1} & m_{2}^{1} & \ldots & m_{K}^{1} \\
\mathbb{P}^{n_2} & m_{1}^{2} & m_{2}^{2} & \ldots & m_{K}^{2} \\
\vdots & \vdots & \vdots & \ddots & \vdots \\
\mathbb{P}^{n_p} & m_{1}^{p} & m_{2}^{p} & \ldots & m_{K}^{p} \\
\end{array}\right]\; .
\end{equation}
Note that the $j^{\rm th}$ column of this matrix contains the multi-degree of the polynomial $p_j$.
In order for the resulting manifold be Calabi-Yau, the condition $\sum_{j=1}^L m^{r}_{j} = n_r + 1 ,\qquad \forall r=1, \ldots, p$ must be imposed, similarly to \eqref{cy_cond} so that $c_1(X)$ vanishes. There are over $7000$ such CY 3-folds that can be built in this way \cite{Candelas:1987kf} (see also \cite{Gray:2013mja} for similar constructions of CY 4-folds).

\begin{figure}[h]
\centering
\includegraphics[width=0.7\textwidth]{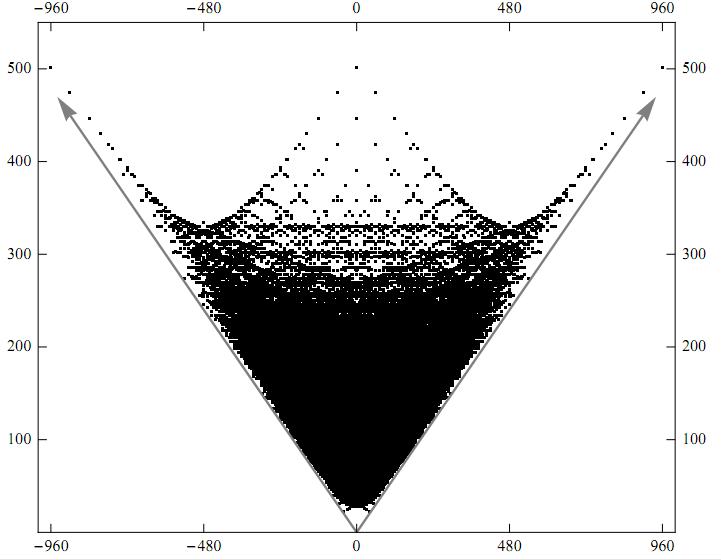}
\caption{\emph{A plot of Hodge numbers arising within the Kreuzer-Skarke dataset of Calabi-Yau threefolds (i.e. anti-canonical hypersurfaces in toric varieties \cite{Kreuzer:2000xy}). The horizontal axis denotes $(h^{1,1}-h^{2,1})$ while the vertical axis is $(h^{1,1}+h^{2,1})$. }}\label{Kreuzer_Skarke}
\end{figure}

By generalizing the ambient spaces still further to more general toric varieties, one can build CY hypersurfaces or more general complete intersections. The most famous dataset of such manifolds is known as the Kreuzer-Skarke dataset \cite{Kreuzer:2000xy} and consists of more than half a billion CY threefolds. The topology of known CY manifolds can vary widely. For example the independent Hodge numbers described in Section \ref{sec:moduli} can range from $(h^{1,1},h^{2,1})=(1,1)$ to $(h^{1,1},h^{2,1})=(11,492)$. A representation of the Hodge numbers found in the Kreuzer-Skarke dataset are shown in Figure \ref{Kreuzer_Skarke}.

\subsection{The holomorphic top form in Complete Intersection Calabi-Yau manifolds}
In this section we will introduce a crucial tool, the form of the holomorphic $3$-form, $\Omega$ in a complete intersection Calabi-Yau (CICY) threefold. This explicit form will prove crucial in numerically implementing the Monge-Ampere equations of \eqref{monge_ampere} as we will see in later lectures.

On a Calabi-Yau threefold, the $(3,0)$ form, $\Omega$, is essentially unique. As a result, in a given construction of the manifold, it may be possible to directly ``spot" a covariantly constant, nowhere-vanishing form. Once found, this can be yield new insights. The formulation I'll review is well-known \cite{Strominger:1985it,Witten:1985xc,Candelas:1987se,Candelas:1987kf} for the CICY threefold dataset here \cite{Candelas:1987kf}. We'll explore this first for the case of a single hypersurface in $\mathbb{P}^n$.

Claim: Let $X$ be given by $p(z)=0$ in $\mathbb{P}^n$. Suppose we work on a coordinate patch with $z^i=1$ in ambient (homogeneous) coordinates. Then
\begin{equation}\label{omega_guess}
\Omega=\int_\gamma \frac{\mu}{p}
\end{equation}
with 
\begin{equation}
\mu=\sum_{j=0}^n (-1)^j \omega_j z^j dz^0 \wedge \ldots \wedge \hat{dz^i} \wedge \ldots \wedge dz^n
\end{equation}
where the term under the hat symbol is omitted and the contour $\gamma$ surrounds $p=0$ inside $\mathbb{P}^n$. For further details we direct the reader to \cite{Candelas:1987kf}.

Here is an intuitive feel for why this formula is correct. Note that the form $\mu$ scales as $\mu \to \lambda^{n+1} \mu$ under the coordinate scaling of $\mathbb{P}^n$, but there exists a natural object that scales the same way. Namely, the defining equation $p(z)$ itself! As a result, the threeform given in \eqref{omega_guess} is manifestly holomorphic and be shown to be nowhere vanishing. As a result, it is the canonical three form we're after.

\subsection{Appendix to Lecture 1: Some Useful Pieces}
We catalog here a few useful results that will be important leading into later sections. 

\begin{Definition}
(The Trace Lemma)
\begin{equation}
\partial \log (\det(A))=\text{tr}(A^{-1}\partial A)
\end{equation}
\end{Definition}

and 
\begin{theorem}
(The Global $\partial\bar{\partial}$-lemma)

\noindent Let $M$ be a K\"ahler manifold and $\phi \in H^{1,1}(M)$ ($\cap H^2(M, \mathbb{R})$) then $\phi$ is exact if and only if there exists a function $f \in C^{\infty}(M,\mathbb{R})$ such that $\phi=\sqrt{-1}\partial \bar{\partial}f$.
\end{theorem}




\section{Lectures 2 \& 3: Ricci-flat metrics from conventional numerical techniques}

In this section, we will review two of the conventional techniques that have historically been used to obtain numerical approximations to Ricci-flat metrics on Calabi-Yau manifolds (for an early example of a different approach see \cite{Headrick:2005ch}). These approaches, while distinct do share some common features. In particular they both follow a two step process.
\begin{itemize}
\item An ansatz is made for the metric. More precisely, since these are K\"ahler metrics, an ansatz is made for the K\"ahler potential, a double derivative of which then returns that metric. This ansatz contains some parameters, which we will denote as $h$ in what follows.
\item A method is then used to adjust the parameters $h$ to get as close as possible to the Ricci-flat metric. How close a metric is to this desired result is measured by an error measure, a functional of the metric which vanishes only for the Ricci-flat instance.
\end{itemize}
It is useful to go through both of these steps in some detail here because much of the setup that is required in doing so is exactly what is required for the Machine Learning application that will be discussed in later sections. 

In the following discussion we will stick, for simplicity, to Calabi-Yau manifolds which can be described in the form $\left[ \mathbb{P}^N|N+1 \right]$. So for example we could consider,
\begin{eqnarray} \nonumber
\text{K3} &\;\;\;& \left[\mathbb{P}^3|4 \right] \\  \label{mans}
\text{the quintic threefold}&\;\;\;& \left[\mathbb{P}^4|5 \right] \\ \nonumber
\text{of the sextic fourfold}&\;\;\;& \left[\mathbb{P}^5|6 \right] \;.
\end{eqnarray}
Much of this section follows standard discussions in the literature and the author found these references \cite{Headrick:2009jz,Douglas:2006rr} particularly useful in compiling it\footnote{In addition, some very useful notes can be found here: \url{https://people.brandeis.edu/~headrick/Mathematica/strategy.pdf}
}.

\subsection{Metric Ansatzs}

Let us define the homogeneous coordinate on $\mathbb{P}^N$ to be $Z^a$ where $a=1,\ldots, N+1$, with the usual associated open patches $Z^c \neq 0$ being denoted by $U_c$. We can then define the quantities $u^a_c= Z^a/Z^c$ which for $a\neq c$ are the usual affine coordinates on the patch $U_c$. The K\"ahler potential for the Fubini-Study metric on $\mathbb{P}^N$ can be written, on patch $U_c$, in terms of these quantities as follows.
\begin{eqnarray} 
\label{fs}
K^{FS}_c := \ln \left( \frac{G_{a \overline{b}} Z^a Z^{\overline{b}}}{|Z^c|^2}\right) = \ln \left(G_{a\overline{b}} u^a_c \overline{u^b_c} \right)
\end{eqnarray}
In this expression, $G_{a \overline{b}}$ is a constant hermitian matrix, and one may obtain the metric via $g^{FS}_{a \overline{b}}= \partial_a \partial_{\overline{b}}K_c^{FS}$ where the derivatives are with respect to the affine coordinates. 

An initial question a reader might have, given such an explicit metric on $\mathbb{P}^N$ is can we not just pull this back to get a metric on the Calabi-Yau manifold itself? More precisely, if we we define some coordinates $y^i$ on the open patches of the Calabi-Yau manifold obtained by intersection with the $U_c$ then we have the following.
\begin{eqnarray}
g^{\text{CY}}_{i \overline{j}} = \frac{\partial u^a_c}{\partial y^i} \frac{\partial \overline{u}^{\overline{b}}_c}{\partial \overline{y}^{{\overline j}}} g^{\text{FS}}_{a \overline{b}}
\end{eqnarray}
The metric $g^{\text{CY}}$ is indeed perfectly well defined on the Calabi-Yau manifold, it is simply not the Ricci flat metric (see \cite{Font:2005td} for an example of a concrete computation). Given this, our goal is to generalize $g^{\text{FS}}$ to something more flexible so we can obtain the Ricci flat metric (or at least a good approximation to it) under the above pull back under the embedding map $y^i \to u^a_c(y^j)$ .

One possibility would be to write down an ansatz for a K\"ahler potential that is given as an expansion in the following objects, which span the first $k$ eigenspaces of the Laplacian on $\mathbb{P}^N$.
\begin{eqnarray}
\frac{P_k^I(Z) \overline{P}^{\overline{J}}_k(\overline{Z})}{\left(G_{a \overline{b}} Z^a \overline{Z}^{\overline{b}} \right)^k}
\end{eqnarray}
One problem with this is that these functions are linearly dependent on $X$. In particular,
\begin{eqnarray} \label{redun}
P_k^I(Z) \sim P_k^I(Z) + \alpha_{k-N-1} P^{\text{defining}}_{N+1}\;,
\end{eqnarray}
where $\alpha$ is an arbitrary polynomial of the indicated degree and $P^{\text{defining}}_{N+1}$ is the defining polynomial of $X$. If we remove this redundancy and only include one representative of each equivalence class, which we shall denote by $p^A_k$, then we can write the following \cite{Tian}. 

\begin{eqnarray}\nonumber
K_{(c)} &=& K_{(c)}^{\text{FS}} + \frac{1}{k} \ln \left( \frac{h_{A \overline{B}} p^A_k \overline{p}^{\overline{B}}_k}{( G_{a \overline{b}} Z^a \overline{Z}^{\overline{b}})^k}\right) \\ \nonumber
&=& \frac{1}{k} \ln \left( h_{A\overline{B}} \frac{p_k^A \overline{p}^{\overline{B}}_k}{|Z^c|^{2k}}\right) \\ \label{algkp}
&=& \frac{1}{k} \ln \left( h_{A\overline{B}} p_k^A(u_{(c)}) \overline{p}^{\overline{B}}_k(\overline{u}_{(c)}) \right)
\end{eqnarray}
In this expression $h$ is an hermitian matrix of parameters (in order to return a real K\"ahler potential). The idea is to take this ansatz and vary the $h$ to try and obtain a metric which, when pulled back to $X$, is a good approximation to the Ricci Flat metric on that manifold.

\vspace{0.1cm}

One question the reader might have is how likely  is it that this attempt to approximate the Ricci Flat metric will succeed. The answer is furnished to us by the following theorem.

\vspace{0.2cm}

\noindent
{\bf Theorem (Tian):} {\it Let $\{p_k^A\}$ be a basis for $H^0(X,{\cal L}^k)$ for some ample line bundle ${\cal L}$. Then the space of all algebraic K\"ahler potentials of the form (\ref{algkp}) where $k\in \mathbb{Z}$ is dense in the space of K\"ahler potentials}

\vspace{0.2cm}
For the case in hand ${\cal L}={\cal O}(1)$ so that our $p_k \in H^0(X, {\cal O}(k)) = H^0(X,{\cal L}^k)$. Thus we see that if we use high enough polynomial degrees in forming our ansatz for $K_{(c)}$ then we can get arbitrarily close to any K\"ahler metric we desire, in particular the Ricci-flat one. Essentially this result makes rigorous the idea that as we raise the polynomial degree $k$ there are more parameters appearing in our metric ansatz, and thus more freedom to tune ourselves closer to the desired metric.

\vspace{0.2cm}

The above discussion leaves something to be desired in terms of geometric intuition as to the nature of the ansatz (\ref{algkp}). In fact, it has a rather simple interpretation. Define a map $\mathfrak{i}: X \to\mathbb{P}^{N_{p_k} -1}$, where $N_{p_k}= h^0(X,{\cal O}(k))$ the number of independent $p_k$, as follows.
\begin{eqnarray} \label{embedding}
\mathfrak{i}(Z^0,\ldots, Z^N) = (X_0=p_k^1(Z),\ldots, X_{N_{p_k}}=p_k^{N_{p_k}}(Z))
\end{eqnarray}
Here the $X_i$ are the homogeneous coordinates of $\mathbb{P}^{N_{p_k} -1}$. Note that this map respects the scaling of the projective spaces correctly as a change $Z \to \lambda Z$ induces a change $p_k \to \lambda^k p_k$. Note also that in order to be a well defined map it must be the case that not all of the $p_k$ can vanish simultaneously. This will certainly be the case for all of the examples discussed in this chapter.

To obtain a nice geometric interpretation for our algebraic ansatz from this map it is useful for it to be an embedding rather than simply an immersion. Mathematically, we will obtain an embedding if $L$ is very ample where,
\begin{eqnarray}
p_k^A \in H^0(X, L)\;.
\end{eqnarray}
This condition holds for the quintic Calabi-Yau threefold in (\ref{mans}), for example, for all $k\geq 1$. 

With these caveats aside what is the geometric interpretation of the algebraic ansatz (\ref{algkp})? Well, given (\ref{embedding}) we see that the K\"ahler potential (\ref{algkp}) is simply the Fubini-Study K\"ahler potential (\ref{fs}) on $\mathbb{P}^{N_{p_k} -1}$ restricted to $X$. Thus the associated metric is simply the pullback of the Fubini-Study metric on that large projective space.

\vspace{0.1cm}

How many free parameters does the K\"ahler potential ansatz (\ref{algkp}) contain? Well, the number of degree $k$ polynomials in the homogeneous coordinates of $\mathbb{P}^N$ is
\begin{eqnarray}
N_k= \left( \frac{k+N}{k}\right)\;.
\end{eqnarray}
Therefore, after removing the redundancy (\ref{redun}), we find the following.
\begin{eqnarray}
N_{p_k} = \left\{\begin{array}{ccc} \left( \frac{k+N}{k}\right)& & \text{If}\; k\leq N  \\  \left( \frac{k+N}{k}\right) - N_{k-(N+1)}&=\left( \frac{k+N}{k}\right)-\left( \frac{k-1}{k-(N+1)}\right)&\text{If}\; k > N\end{array}\right.
\end{eqnarray}
This expression scales as $k^{N-1}$ with the degree and dimension of the projective ambient space and thus increases quickly with degree. For the quintic, for example, the expression reduces to $\frac{5}{6} k(5+k^2)$ leading to the following values.

\begin{center}
\begin{tabular} {c|ccccc} 
k & 5&6&7&8&9 \\  \hline
\# polynomials $p_k$ & 125&205&315&460&645
\end{tabular}
\end{center}

It should be remembered that the parameters $h$ form a matrix with each index running over these possibiities. Thus we arrive at the following for the number of parameters in the algebraic ansatz.

\begin{center}
\begin{tabular} {c|ccccc} 
k & 5&6&7&8&9 \\  \hline
\# parameters $h_{A\overline{B}}$ & 15625&42025&99225&211600&416025
\end{tabular}
\end{center}

In almost all conventional applications (not employing Machine Learning techniques) to finding Ricci-flat metrics, the above numbers of parameters are simply too large to deal with. Thus, some simplification must be employed to decrease the number of degrees of freedom involved.

The most common method for decreasing the number of parameters that must be considered is to impose a symmetry on the Calabi-Yau manifold \cite{Headrick:2009jz,Douglas:2006rr,Braun:2010vc,Gray:2021kax}. We chose a special defining relation such that such a symmetry exists and then restrict to the space of polynomials invariant under that symmetry. For example, consider the following defining relation for the quintic.
\begin{eqnarray}
P_5^{\text{defining}} = \sum_{a=0}^4 (Z^a)^5
\end{eqnarray}
For this choice we have the following symmetries which leave the defining relation invariant \cite{Headrick:2009jz,Douglas:2006rr}.
\begin{itemize}
\item Permutation of the $Z^a$.
\item The action $Z^a \to \omega^a Z^a$ where the $\omega^a$ are any fifth roots of unity (which can be different for different $a$'s).
\item Complex conjugation of the coordinates.
\end{itemize}
These action give rise to a discrete symmetry group of order 150,000. Imposing this symmetry reduces the number of parameters we must consider in the algebraic K\"ahler potential to ${\cal O}(10)$ at $k=9$, to be compared with $416025$ in the above table \cite{Headrick:2009jz}. It should be emphasized that the ability to avoid having to impose such symmetries is one of the areas, along with moduli dependence, where Machine Learning techniques have made qualitative improvements to the results which can be obtained.

\vspace{0.2cm}

As one last comment about the algebraic K\"ahler metric, let us talk about some possible choices of coordinates on the Calabi-Yau manifold and how to compute the necessary derivatives to obtain the metric (and it's derivatives) efficiently. Locally, we could use any subset of $N-1$ out of our $N$ affine coordinates $u_c^a$ on the patch $U_c$ of $\mathbb{P}^n$ as coordinates on $X$. Following the notation of \cite{Headrick:2009jz}, we can pick any $\delta \neq c$ and our Calabi-Yau coordinates are then,
\begin{eqnarray}
z^i = u^i_c \;\; (i\neq c,\delta)\;.
\end{eqnarray}
In such a parameterization $u^{\delta}_c$ is a function of the $z^i$ obtained by solving $P_5^{\text{defining}}=0$.

Naively, taking a derivative  with respect to these coordinates is painful, as one must use the chain rule in an appropriate fashion and then substitute in an appropriate solution to the defining relation for $u^{\delta}_c$.  However, if dealt with correctly, such an arduous procedure can be avoided. Imagine taking the derivative with respect to the Calabi-Yau coordinates of some function $f$. In what follows we shall use an index $\alpha: a| a\neq c$.
\begin{eqnarray} \label{fderiv}
\frac{\partial}{\partial z^i} f(u^{\alpha}_c(z^i)) &=& \frac{\partial f}{\partial u^{\alpha_c}} \frac{\partial u^{\alpha}}{\partial z^i} \\ \nonumber
&=& \frac{\partial f}{\partial u^i_c} + \frac{\partial f}{\partial u^{\delta}_c} \frac{\partial u^{\delta}{_c}}{\partial u^i_c}
\end{eqnarray}
However, the defining relation is constant (vanishing) everywhere on the Calabi-Yau manifold, so that
\begin{eqnarray}
\frac{\partial P^{\text{defining}}}{\partial z^i}= \frac{\partial P^{\text{defining}}}{\partial u^i_c} + \frac{\partial P^{\text{defining}}}{\partial u^{\delta}_c} \frac{\partial u^{\delta}_c}{\partial u^i_c}=0 \;.
\end{eqnarray}
We therefore find that,
\begin{eqnarray}
\frac{\partial u^{\delta}_c}{\partial u^i_c} = -\frac{\partial P^{\text{defining}}}{\partial u^{i}_c}/ \frac{\partial P^{\text{defining}}}{\partial u^{\delta}_c} =- P_i^{\text{def}}/P_{\delta}^{\text{def}}
\end{eqnarray}
Where we use the last equality to define $P_i^{\text{def}}$ and $P_{\delta}^{\text{def}}$. Using this result we can write the following expression for  (\ref{fderiv}).
\begin{eqnarray}
\frac{\partial f}{\partial z^i} = \frac{\partial f}{\partial u^i_c} - \frac{P_i^{\text{def}}}{P_{\delta}^{\text{def}}} \frac{\partial f}{\partial u^{\delta}_c}
\end{eqnarray}
This type of expression allows us to easily compute derivatives with respect to Calabi-Yau coordinates in terms of ambient space quantities. As described above, this is necessary for example to obtain the metric on the Calabi-Yau manifold itself.

\subsection{Tuning the parameters}

Given the algebraic ansatz for the K\"ahler potential discussed in the last subsection, how will we fix the parameters in the ansatz to obtain Ricci-flat metrics? There were two commonly used techniques to be found in the literature before the advent of Machine Learning methodologies.
\begin{itemize}
\item The Donaldson algorithm: the iteration of a certain operator to a fixed point (see \cite{Douglas:2006rr} for an early application in this context).
\item The minimization of an appropriate functional \cite{Headrick:2009jz}.
\end{itemize}
We will start with the latter possibility first here, as it is the method that generalizes most naturally to the context of the Machine Learning techniques discussed in the next section.

\subsubsection{Some basic definitions}

Let us start by defining some of the quantities that we will need in the subsequent discussion. If we denote the complex dimension of $X$ by $n=N-1$ then we define the following top form.
\begin{eqnarray}
\mu= (-i)^n \Omega \wedge \overline{\Omega}
\end{eqnarray}
The factor of $i$ in that expression is included such that $\mu$ is real. Given a K\"ahler form $J$ we can then define the following scalar quantity.
\begin{eqnarray} \label{vdef}
v= \frac{J^n}{n! \mu}
\end{eqnarray}
Since both the numerator and the denominator in the above are top forms they are proportional to each other and so the ratio $v$ is well defined. Direct computation then reveals the following explicit formula.
\begin{eqnarray} \label{vexp}
v =\frac{\det g_{i\overline{j}}}{|\Omega_{1\ldots n}|^2}
\end{eqnarray}
A useful property of the quantity thus defined is that the equation $v=\text{constant}$ (call it $c$) becomes the following.
\begin{eqnarray}
\frac{1}{n!} J^n = c \Omega \wedge \overline{\Omega} (-i)^n
\end{eqnarray}
This is precisely the Monge Ampere equation. Finally, there is a simple relationship between $v$ and the Ricci tensor for the metric associated to $J$.
\begin{eqnarray} \label{riccifromv}
R_{i \overline{j}} = - \partial_{i} \partial_{\overline{j}} \ln v
\end{eqnarray}
It is useful to note in deriving this expression that, from (\ref{vexp}),
\begin{eqnarray}
\ln v= \ln \det g_{i\overline{j}} - \ln \Omega_{1\ldots n}- \ln \overline{\Omega}_{\overline{1}\ldots\overline{n}}\;.
\end{eqnarray}

\subsubsection{Energy functionals}

In forming functionals whose minimization would lead to a Ricci flat metric on $X$ we might want something that is minimized either by solving the Monge Ampere equation or by the Ricci flatness condition \cite{Headrick:2009jz}. Lets start with the first of these, which is often called the Monge Ampere Functional.
\begin{eqnarray} \label{eq:MAfunc}
H_{\text{MA}} := \int_X \mu (v- <v>)^2
\end{eqnarray}
Here,
\begin{eqnarray}
<v> := \frac{\int_X \mu v}{\int_x \mu}\;,
\end{eqnarray}
is simply the average of $v$ over the Calabi-Yau manifold, computed using the measure $\mu$. Through $v$'s definition (\ref{vdef}) this is a functional of $J$ and thus, if evaluated with the ansatz (\ref{algkp}) a function of the parameters $h$.

The functional $H_{\text{MA}}$ has several desirable properties. It is positive semi-definite. The functional has a global minimum at $v= <v>$ which corresponds to,
\begin{eqnarray} 
\frac{J^n}{n!} = <v> \Omega \wedge \overline{\Omega} (-i)^n\;.
\end{eqnarray}
In other words, its global minimum corresponds to solutions to the Monge Ampere equation. The functional has no other critical points as can be shown by direct computation. Lastly $H_{\text{MA}}$ can be computed without taking any derivatives of $J$. Since $\mu$ is fixed and known in closed form for the types of Calabi-Yau we are considering here \cite{Candelas:1987kf} this makes the use of this functional pretty efficient and simple!

\vspace{0.2cm}

Next let us consider what we shall call the Ricci functional.
\begin{eqnarray} \label{Rfuncß}
H_R  &&:= -\frac{1}{2} \int_X \mu R \\\nonumber
&&= \int_X \mu g^{i \overline{j}} \partial_i \partial_{\overline{j}} \ln v
\end{eqnarray}
Here $R$ is the Ricci scalar and the second line follows from (\ref{riccifromv}). We can rewrite this functional to make some of its properties clearer. Recall that, from (\ref{vdef})
\begin{eqnarray}
\mu = \frac{J^n}{n! v} = \frac{\det g_{i\overline{j}}}{n! v} \epsilon^{(s)}\;,
\end{eqnarray}
where $\epsilon^{(s)}$ is the anti-symmetric symbol and $\det g_{i \overline{j}} = \sqrt{g}$. Then we can write the following.
\begin{eqnarray}
H_R &=&\int_X \frac{\sqrt{g}}{n! v} \epsilon^{(s)}g^{i\overline{j}} \partial_i \partial_{\overline{j}} \ln v \\ \nonumber
&=& - \int_X \partial_{\overline{j}} \ln v \partial_i \left( g^{i\overline{j}} \sqrt{g} \frac{\epsilon^{(s)}}{n!v}\right) \\ \nonumber
&=& - \int_X \epsilon^{(s)} \frac{\sqrt{g}}{n!}( \partial_{\overline{j}} \ln v )g^{i\overline{j}} (\nabla_i \frac{1}{v} )\\ \nonumber
&=& \int_X \mu g^{i\overline{j}} (\partial_i \ln v )(\partial_{\overline{j}} \ln v)
\end{eqnarray}
With the functional in this form we can now see that it enjoys several useful properties. It is positive semi-definite. It has a global minimum when $v=\text{constant}$, which as we have already established corresponds to the Ricci-flatness condition. The functional has no other critical points beyond these global minima, as can be determined by direct computation. However, this functional is a bit uglier and more difficult to use than $H_{\text{MA}}$. In particular it's use requires taking derivatives of $J$ which often leads to it being a less efficient choice.

\subsubsection{Integration and point sampling}

To evaluate the functionals presented in the last subsection, and indeed to use the Donaldson algorithm that we will present shortly, we need to be able to perform integrals over the Calabi-Yau manifold $X$. For the applications at hand these integrals need to be carried out numerically and so we briefly review some appropriate techniques here \cite{Douglas:2006rr,numrec}.

Let us start by considering the integration of a simple object where we can spot the answer. Consider an open set $U\subset X$. We can then define,
\begin{eqnarray}
1\!\!1_U(x) = \left\{ \begin{array}{ccc} 1 & \text{if} & x\in U \\ 0 & \text{if} & x \notin U\end{array}\right.\;.
\end{eqnarray}
It is then intuitively clear that,
\begin{eqnarray}
\int_X 1\!\!1_U \mu = \text{Vol}_{\mu}(U)\;,
\end{eqnarray}
where $\text{Vol}_{\mu}(U)$ is the volume of $U$ {\it with respect to the measure} $\mu$.

{\it If} we have a sample of $N_p$ points $\{q_i \in X\}$ on $X$ which is {\it uniformly distributed according to} $\mu$ then the expected number of points in $U\subset X$ is, 
\begin{eqnarray}
\sum_{i=1}^{N_p} 1\!\!1_U(q_i) \approx N_p \frac{\text{Vol}_{\mu}(U)}{\text{Vol}_{\mu} X} \;.
\end{eqnarray}
Rearranging this expression, we arrive at an approximate formula for the integral of $1\!\!1_U$.
\begin{eqnarray}
\int_X 1\!\!1_U \mu \approx \frac{\text{Vol}_{\mu}(X)}{N_p} \sum_{i=1}^{N_p} 1\!\!1_U(q_i)
\end{eqnarray}
If instead we wanted to integrate a different, arbitrary function, one might expect that one gets a similar result, but with the points being weighted by the value of that function rather than by the $1/0$ weighting afforded by $1\!\!1_U$. This is indeed the case and we arrive at the following approximate formula for the integral.
\begin{eqnarray} \label{fint}
\int_X f \mu \approx \frac{\text{Vol}_{\mu}(X)}{N_p} \sum_{i=1}^{N_p} f(q_i)
\end{eqnarray}
This formula is rather easy to understand. The prefactor to the sum assigns a volume to each point in the sample (which is the same for every point due to the ``uniformly distributed" assumption). One then simply multiplies this volume by the value of $f$ at each point and sums  to get the approximation to the integral. The statistical error in such an approximation is of order $1/\sqrt{N_p}$ \cite{numrec}.

\vspace{0.2cm}

In practice, the above approach to performing integrals over a Calabi-Yau manifold $X$ encounters a problem in that producing samples of points distributed according to $\mu$ is not easy! Instead, one modifies the above approach in a two step process.
\begin{itemize}
\item First, one produces a set of points uniformly distributed according to some other measure $\mu'$ which is well adapted to such a task.
\item Second, one then follows the above procedure but accounts for the fact that the points were not distributed according to $\mu$ by making use of what are called `mass functions'. 
\end{itemize}

Specifically, we define a mass function associated to some measure $\mu'$ by,
\begin{eqnarray}
m(x) = \frac{\mu(x)}{\mu'(x)}\;.
\end{eqnarray}
As in our definition of $v$ above, such a ratio is well defined as $\mu$ and $\mu'$ are both top forms and are thus proportional to one another. We then have the following for an approximation to the integral of some function $f$. 
\begin{eqnarray} \label{preintmass}
\int_x f \mu = \int_X f \frac{\mu}{\mu'} \mu'  \approx \frac{\text{Vol}_{\mu'}}{N_p} \sum_{i=1}^{N_p} f(q_i) m(q_i)
\end{eqnarray}
Here in the approximate equality we have used our earlier result (\ref{fint}). One can process this result further by noting that because,
\begin{eqnarray}
\int_X \mu = \text{Vol}_{\mu}(X) = \int_X \frac{\mu}{\mu'} \mu' \approx \frac{\text{Vol}_{\mu'}(X)}{N_p} \sum_{i=1}^{N_p} m(q_i) \;,
\end{eqnarray}
we have the following result.
\begin{eqnarray}
\frac{\text{Vol}_{\mu'}(X)}{N_p} \approx \frac{\text{Vol}_{\mu}(X)}{\sum_{i=1}^{N_p} m(q_i)}
\end{eqnarray}
Substituting this result into (\ref{preintmass}) we arrive at the following expression \cite{Douglas:2006rr}.
\begin{eqnarray}
\int_X f \mu \approx \frac{\text{Vol}_{\mu}(X)}{\sum_{i=1}^{N_p} m(q_i)} \sum_{i=1}^{N_p} f(q_i) m(q_i)
\end{eqnarray}
The advantage of this expression is that it makes no mention of the auxiliary measure $\mu'$ that was used to form the point sample, outside of the mass function $m$.

\vspace{0.2cm}

What kind of point sampling techniques are used in the literature on approximate Ricci flat metrics on Calabi-Yau manifolds? Here we will give two of the most common examples.

\subsubsection*{Rejection sampling}

\begin{figure}[t!]
\begin{center}
\includegraphics[scale=0.5]{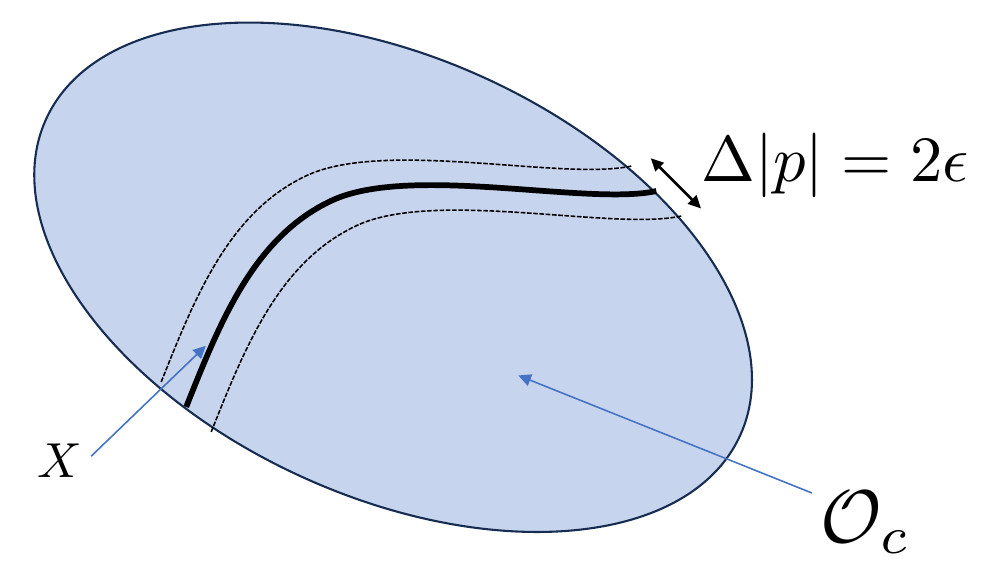}
\end{center}
  \caption{\label{rejsamp}{\it A schematic depiction of some quantities appearing in rejection sampling, as described in the text.}}
  \end{figure}

The first method we will describe was used, for example, in \cite{Headrick:2009jz}. Begin by defining a cover of the manifold $X$ in terms of sets of points  ${\cal O}_c$ where $\left|\frac{Z^{\alpha}}{Z^c}\right| \leq 1 \;\; \forall \;\alpha \neq c$. These coordinate patches are useful in numerical work because the affine coordinates associated to them are finite in value (less than or equal to one in magnitude). One then carries out the following steps.
\begin{itemize}
\item Choose points randomly in the patches ${\cal O}_c$ according to the coordinate measure.
\item Throw away any points for which $|P^{\text{defining}}_{N+1}|\geq \epsilon$, where $\epsilon$ is some small number. In other words, we are only keeping those points which are already close to being on the Calabi-Yau. This situation is depicted schematically in Figure \ref{rejsamp}.
\item Project the remaining points onto $X$ orthogonally with respect to the Fubini-Study metric to obtain the sample.
\end{itemize}
  
This point sampling method has a number of issues. First, it is wasteful in that one generates a large number of points far from the Calabi-Yau manifold that are then immediately discarded.  This makes the method somewhat slow compared to other possibilities. Second, it is hard to know the measure $\mu'$ with respect to which the points which are generated in this fashion are distributed. Still, the method is fast enough for most purposes and in some respects the second issue mentioned above doesn't really matter so much. If we were to ignore the mass functions in this situation and just compute the integral then we are really computing the integral
\begin{eqnarray}
\int_X \mu' (v-<v>)^2
\end{eqnarray}
instead of 
\begin{eqnarray}
\int_X \mu (v-<v>)^2\;.
\end{eqnarray}
Both $\mu$ and $\mu'$ are just some fixed measures so in some respects both of these functionals are valid as quantities to minimize in order to find Ricci-flat metrics. That said, $\mu$ does have the advantage that it is proportional to $J\wedge J \wedge J$, the metric volume form, upon extremization of the functional. Thus in situations where, for example, there are large physical volume regions associated to small coordinate volume regions on $X$ one can expect a better approximation from using $\mu$ than $\mu'$. Nevertheless, in many situations this subtlety is not important, and good results can be obtained with this simple point sampling technique \cite{Headrick:2009jz}.

\subsubsection*{Homogeneous sampling in projective space}

\begin{figure}[t!]
\begin{center}
\includegraphics[scale=0.5]{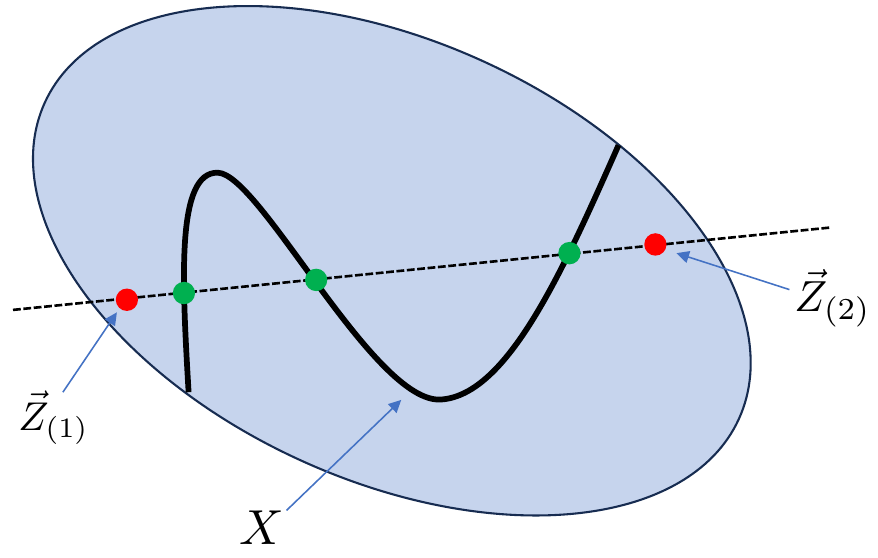}
\end{center}
  \caption{\label{homsamp}{\it A schematic depiction of some quantities appearing in the point sampling method based upon homogeneous sampling in projective space, as described in the text.}}
  \end{figure}

The second method we will describe for point sampling was used, for example, in \cite{Douglas:2006rr}. The method is based upon starting with a uniform sample of pairs of points $\vec{Z}_{(1)}$ and $\vec{Z}_{(2)} \in \mathbb{P}^N$. Such pairs are easy to obtain. For example one can perform the following two steps to find individual points on $\mathbb{P}^N$.
\begin{itemize}
\item Generate uniformly distributed points on $S^{2N+1}$. This can be achieved by using the Gaussian distribution (or any other desired spherically symmetric distribution)
\begin{eqnarray}
\frac{1}{\sigma \sqrt{2 \pi}} e^{-\frac{1}{2} \frac{1}{\sigma^2} \sum_a |Z^a|^2}
\end{eqnarray}
and dividing by the norm.
\item One can then obtain points on $\mathbb{P}^N$ by quotienting out by the $U(1)$ action $Z^i \to e^{i \theta} Z^i$.
\end{itemize}
Given pairs of such points one can then proceed as follows to obtain a sample of points on $X$ itself.
\begin{itemize}
\item Select pairs of points $\{\vec{Z}_{(1)},\vec{Z}_{(2)}\}$ in projective space as just described.
\item Compute
\begin{eqnarray}
P^{\text{defining}}_{N+1} (\vec{Z}_{(1)} + t \vec{Z}_{(2)}) =0
\end{eqnarray}
and solve this to obtain all possible values for $t$. This finds for us the homogeneous coordinates, $\vec{Z}_{(1)} + t \vec{Z}_{(2)}$ with those values of $t$ substituted in, of the $N+1$ points on the Calabi-Yau manifold $X$ which are the intersections of that hypersurface with the straight line going through the points with coordinates $\vec{Z}_{(1)}$ and $\vec{Z}_{(2)}$. This situation is depicted in Figure \ref{homsamp}.
\end{itemize}

This method of point sampling has a number of advantages. First, it is not as wasteful as rejection sampling, leading it to be an order of magnitude or so faster in practical implementations. More importantly, perhaps, the resulting points that are obtained in this way on $X$ are distributed with respect to a {\bf known} measure, the Fubini-Study one (see \cite{Douglas:2006rr} for a proof of this using original results by Shiffman and Zelditch \cite{sz1,sz2}). This means we can explicitly compute mass functions in this case.

\subsubsection{Putting it all together}

Finally we have everything that we need to obtain an approximate Ricci-flat metric on a Calabi-Yau manifold $X$ using the functional minimization approach. Here is the procedure.
\begin{itemize}
\item Choose a polynomial degree $k$ and make a metric ansatz of the form (\ref{algkp}) with associated parameters $h$.
\item Choose a functional you wish to minimize.
\item Use numeric integration to compute the functional as a function of the $h$.
\item Minimize with respect to the parameters $h$. In the examples considered in the literature, in the case where symmetries are imposed on $X$, this step has been sufficiently straightforward that it can be completed using Mathematica's ``FindMinimum" command \cite{Headrick:2009jz}.
\item Plug the minimizing $h$'s back into the ansatz to obtain the approximate Ricci-flat metric.
\end{itemize}

\subsubsection{Another approach to tuning the parameters: the Donaldson algorithm}

Let us end this section by briefly describing another commonly used approach to tuning the parameters $h$ in the ansatz (\ref{algkp}) to get close to a Ricci-flat metric. To use this method, called Donaldson's algorithm \cite{Douglas:2006rr,Douglas:2006hz,donaldson2005numerical,donaldson2005scalar,Braun:2007sn,Anderson:2010ke,Anderson:2011ed}, we first define a ``T-Operator" as follows.
\begin{eqnarray}
T_k(h)^{A \overline{B}}:= \frac{N_{p_k}}{\int_X \mu} \int_X{ \frac{p_k^A \overline{p}_k^{\overline{B}}}{h_{C\overline{D}} p_k^X \overline{p}_k^{\overline{D}}}\,\mu}
\end{eqnarray}
Given a Calabi-Yau manifold this operator takes in a matrix of parameters $h$ and outputs a new matrix $T$. This operator exhibits the following property.

\vspace{0.2cm}

\noindent
{\bf Theorem (Donaldson):} {\it For any initial $h_{A\overline{B}}$ and fixed $k$, the sequence
\begin{eqnarray}
h_{m+1} = \left(T_k(h_m)\right)^{-1}
\end{eqnarray}
converges to a unique ``balanced" value as $n\to \infty$ and the associated metric on $X$ is called the ``balanced metric".}

\vspace{0.2cm}

In practice this iteration converges in ${\cal O}(10)$ steps for the types of Calabi-Yau manifolds we are considering in this chapter. This is useful because of the following theorem.

\vspace{0.2cm}

\noindent
{\bf Theorem (Donaldson):}
{\it As $k\to \infty$ the balanced metric on $X$ converges to the unique Ricci-flat metric for a given K\"ahler class and complex structure.}

\vspace{0.2cm}

More generally one can consider non-Calabi-Yau examples and the balanced metric converges to a K\"ahler metric associated to the K\"ahler class $c_1({\cal L})$ where the $p_k$ are sections of ${\cal L}$. This will be a constant scalar curvature metric with curvature determined by $c_1(X)$.

\vspace{0.1cm}

A natural question is why we have not focused so much on this method of tuning the parameters in this section. Here are two of the key reasons.
\begin{itemize}
\item This method is not as directly relevant for most of the Machine Learning approaches to finding Ricci-flat metrics as the functional minimization method.
\item This method tends not to do as well as the functional approach. One reason for this is simply that, while it is true that the balanced metric will converge to the Ricci-flat metric as $k\to \infty$, that does {\it not} mean that for any fixed finite $k$ the balanced metric is the best approximation to that desired result. Indeed, it has been shown that this is not the case \cite{Headrick:2009jz}. Thus minimizing a functional tends to result in a more accurate metric than iterating the T-operator, even while using the same metric ansatz with the same degree polynomials in it.
\end{itemize}

\subsection{Performance of and problems with conventional numerical methods}
\label{sec:perf&prob}

In terms of performance, what diagnostics do we have to see how close we came to obtaining the Ricci-flat metric on our Calabi-Yau manifold $X$? There are a plethora of quantities we can use in such an evaluation. These include the value of both of the functionals discussed in this section, the value of $v$ on a random of selection of points on $X$ ($v$ should be constant for the Ricci-flat metric), or the Ricci scalar evaluated on points. How accurate a metric we obtain will of course depend on $k$, the degree of the polynomials we use in the algebraic K\"ahler potential ansatz (\ref{algkp}). To give an idea of the type of accuracy that has been obtained with conventional techniques, in \cite{Headrick:2009jz}, the functional minimization approach was introduced and was used to obtain values of order $H_{\text{MA}} \sim 0.07 \times 5^{-k}$. So we get something of order $10^{-9}$ for $k=12$ or $10^{-6}$ for $k=6$. By contrast, the Donaldson algorithm only achieves the $k=6$ value for $H_{\text{MA}}$ when used at the $k=12$ level, and scales as $H_{\text{MA}} \sim \frac{\alpha}{k^2}+ \frac{\beta}{k^3}$. This is inline with expectations from the above discussion, and it should be noted that performing the Donaldson computation at $k=12$ takes roughly 2 days on a modern desktop machine.

\vspace{0.1cm}

What are the errors in the results that are obtained? In terms of how close we are to the optimal metric for a given $k$ these can be estimated by a boot strap method \cite{Headrick:2009jz}. We compute the metric multiple times and then calculate the standard deviation of the results. In \cite{Headrick:2009jz} this was carried out and it was found that the error was down by a factor of $1/\sqrt{N_{p}}$ from the expected values. In short, these errors are negligible when compared to those resulting from working at finite $k$. However, are the finite $k$ approximations that are obtained good enough? This depends entirely on the application that one requires the Ricci-flat metric for! Not many such applications have been pursued yet in the literature (see for example \cite{Braun:2008jp,Cui:2019uhy}), due to the drawbacks that these conventional numerical techniques suffer from. It is to these that we now turn.

\vspace{0.1cm}

The conventional numerical approaches to finding approximate Ricci-flat metrics on Calabi-Yau manifolds that we have been discussing in this section suffer from a couple of qualitative drawbacks.
\begin{itemize}
\item For the computations to be practically feasible symmetries must be imposed upon the Calabi-Yau manifolds of interest. As described earlier this reduces the number of parameters that need to be considered in our ansatzes for the K\"ahler potential. Unfortunately, in many applications one would like to consider more generic Calabi-Yau manifolds.
\item All of the computations that have been carried out in the literature with conventional numerical techniques are performed at a single point in moduli space. This is a severe limitation for most applications. Often, one wishes to know this metric so that it is possible to compute those parts of the effective theory of a string theory compactification that depend upon its exact form. We want those potential and kinetic terms as a function of the moduli in almost all instances.
\end{itemize}
It is the last of these issues in particular that has lead to very little progress being made with the numerical metrics that we have managed to compute in the past. Fortunately, both of these drawbacks can be addressed by the use of Machine Learning techniques. For example, there is work computing Ricci-flat metrics as a function of complex structure moduli \cite{Anderson:2020hux, Gerdes:2022nzr}. In addition, in almost all of the Machine Learning literature in this subject there is no need to impose any symmetry constraint on the Calabi-Yau manifold. One can also persue more novel approaches in the Machine Learning context, for example describing $g$ directly with a neural network, rather than using a K\"ahler potential ansatz such as (\ref{algkp}). This has allowed us to investigate non-K\"ahler string compactifications as well \cite{Anderson:2020hux}. It is to these Machine Learning based approaches to the problem of finding metrics on compactifications manifolds that we turn in the next section.

 \section{Lectures 4 \& 5: Ricci-flat metrics from neural networks}



By now, it should be evident that the implementations of Donaldson's algorithm \cite{donaldson2005numerical}, and the numerical minimization of Headrick and Nassar \cite{Headrick:2009jz}, have practical limitations. These numerical methods rely on an algebraic ansatz, where the K\"ahler potential is expanded in a spectral basis (homogeneous polynomials of degree $k$). This  has an inherent scaling problem: the number of monomials $N_k$ grows with $k$, and the number of parameters of the expansion grows like $N_k^2$. 
An additional conundrum is that it is computationally non-trivial to restrict the ambient space polynomials to the Calabi-Yau manifold, by quotienting the spectral basis by the ideal of the defining polynomial.\footnote{The restrictions requires a Groebner basis analysis, which is computationally costly also for moderate $k$.} 

At special points in the CY moduli space, where the polynomial basis is constrained by discrete symmetry groups, this scaling problem is circumvented by restricting the expansion of the K\"ahler potential to invariant elements of the basis. Thus, whenever we are interested in the Ricci flat metric as such special points, the methods at hand suffice. This is true for the Fermat and Dwork quintics, and this is the main reason why these CY manifolds have been favoured in the literature on numerical CY metrics. 

However, this means that traditional numerical methods often fail at generic points in moduli space (for a counterexample, see \cite{Braun:2007sn}). Similarily, the methods run in trouble for the vast majority of CY manifolds that are not defined as one hypersurface in one projective ambient space. The CICY manifolds, classified in by Candelas et.al.~\cite{Candelas:1987kf} and the CY hypersurfaces in the toric ambient spaces classified by Kreuzer and Skarke \cite{Kreuzer:2000xy}, provide example geometries where,  for $k\ge2$, the dimension $N_k$ is prohibitively large.

This is the main motivation to employ machine learning methods to approximate the Ricci flat metric $g_{CY}$ on CY manifolds. After all, the metric we seek to approximate is determined by a function, the K\"ahler potential, and the universal approximation theorems state that a sufficiently elaborate neural network can approximate any function.  In these lectures we will review the various approaches that have recently been employed for such experiments \cite{Ashmore:2019wzb,Anderson:2020hux,Jejjala:2020wcc,Douglas:2020hpv,Ashmore:2021ohf,Larfors:2021pbb,Larfors:2022nep,Gerdes:2022nzr}. We will also comment on how such methods can improve on the scope and accuracy for CY metric approximation. One aim is to allow you to use these tools for your own research, so the lecture will include practical experiments and refer to open-source code that you may download and experiment with. To date, there are three packages available for ML of CY metrics, namely  

    \texttt{cymetric}: \url{https://github.com/pythoncymetric/cymetric}
    
    \texttt{MLgeometry}: \url{https://github.com/yidiq7/MLGeometry}
    
    \texttt{cyjax}: \url{https://github.com/ml4physics/cyjax}
    
 The first of these packages, \texttt{cymetric},  currently has the broadest range of applicability, and will therefore be the main focus of these lectures. For the concrete example studied in the next section, however, any of these packages will do.

\subsection{Machine learning on the quintic}


To keep our discussion concrete, let ut set up the problem. We want to find a Ricci flat metric on a Calabi--Yau 3-fold, which we will assume is constructed either as a CICY \cite{Candelas:1987kf} or as a hypersurface in a toric ambient space from Kreuzer and Skarke's list \cite{Kreuzer:2000xy}. For definiteness we may take this manifold to be the quintic, $X$, defined as a hypersurface in $\mathbb{P}^4$ as the zero locus of a quintic polynomial $p(Z)=0$. To be even more specific, we can take the polynomial to be 
\begin{equation}
   p = \sum_{a=0}^4 (Z^a)^5
\end{equation}
where $Z^a$ are the homogeneous coordinates of $\mathbb{P}^4$. This would define the Fermat quintic, arguable the simplest CY 3-fold we can think of. Any other qunitic polynomial would do, and would provide another space in the family of quintics, which is related to the Fermat quintic by a complex structure deformation.\footnote{We could also work in other dimensions, say with a quartic CY 2-fold. As a rule of thumb, reducing the dimension of the CY manifold will speed up ML experiments.}

As we know from the Calabi--Yau theorem, finding the Ricci flat metric (in a given K\"ahler class) is equivalent to solving the Monge-Amp\`ere equation
\begin{equation}
    J \wedge J \wedge J = \kappa \, \Omega \wedge \overline{\Omega}
\end{equation}
where $\kappa$ is a constant.\footnote{This means that $\kappa$ is constant with respect to the $CY$ coordinates; it will in general depend on the K\"ahler and complex structure moduli.} We want to solve this equation for $J$, which defines the CY metric. To our aid we have $\Omega$, which can be computed, patch by patch, as a residue in the ambient space; this was explained in the first lecture of this series, see Section \ref{lect1}. We also recall that $J$ is related to any known representative in the specified cohomology class, such as the Fubini--Study (FS) form $J_{FS}$, by 
\begin{equation}
     J = J_{FS} + \partial \bar{\partial} \phi
\end{equation}
for a function $\phi$. This is helpful, since FS Kähler forms may be constructed by restriction from the corresponding ambient space forms, which, as explained in Section \ref{lect1}, we know how to compute.

\begin{figure}[htb]
    \centering
    \includegraphics[width=
0.7\textwidth]{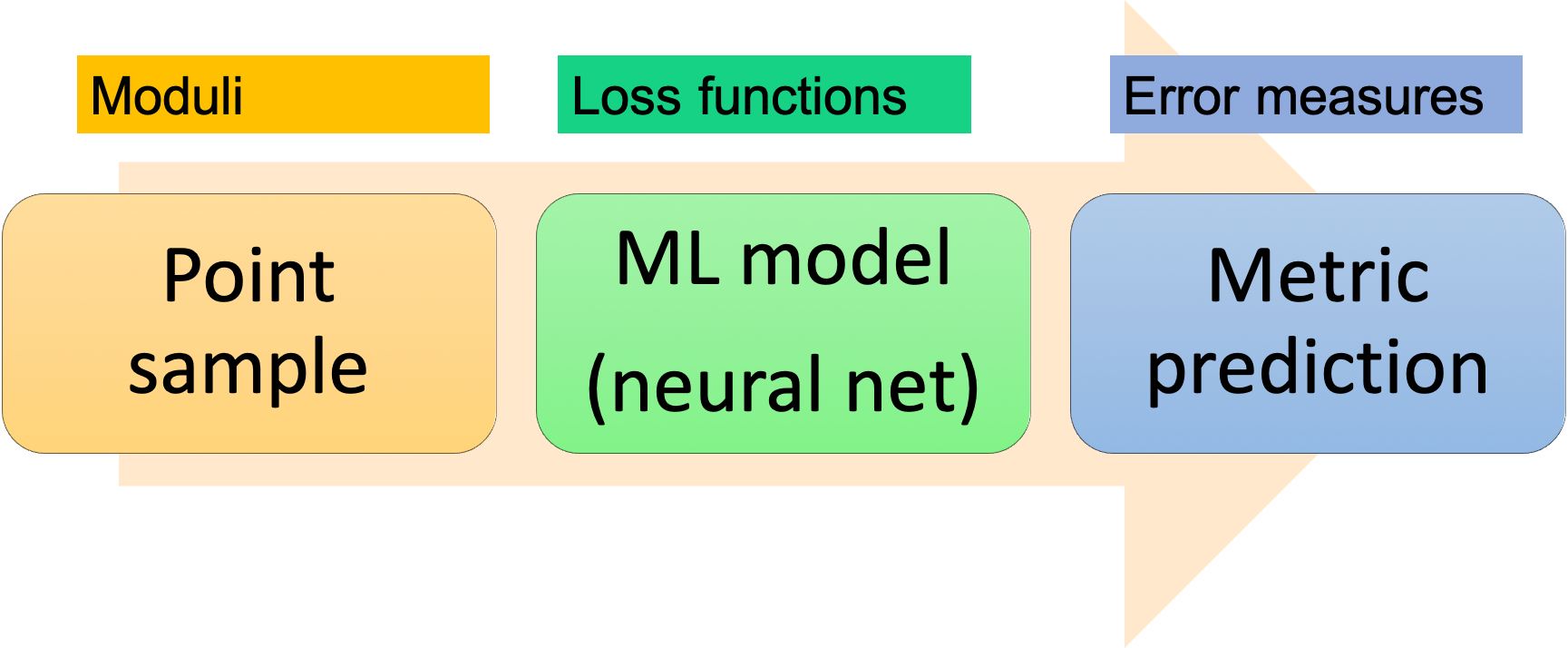}
    \caption{Any ML implementation of CY metric predictions first requires a point sampling routine, for the complex structure and K\"ahler moduli specified by the user. Second, a neural network should be trained, on the point sample, to satisfy the constraints on the metric, which are encoded as loss functions. Finally, the accuracy should be tested using standard error measures.}
    \label{fig:implementation}
\end{figure}

Thus, our machine learning implementation should be such that we can compute $\Omega$ and $J_{FS}$ in the points we sample on the CY 3-fold. We will then feed this information to a neural network that should provide an approximation of the Ricci-flat metric $g$, or the related K\"ahler potential $K$. We will train the network using the constraints we have for the problem, such as the Monge-Amp\`ere equation, which we encode as loss functions to be minimized by during training. Finally, we should check the accuracy of the approximation, for example by computing the Ricci tensor of the predicted metric. Encoding these checks  as error measures, we can trace their evolution during training of the network. We want to do this in a systematic and modular fashion, so that our implementation can be readily adapted to different points in moduli space, and eventually to different CY manifolds. A schematic view of the implementation is presented in Figure \ref{fig:implementation}. 

We see that this machine learning implementation follows the traditional methods to a large degree. First, the same point sampling routines described above for the quintic may be employed\footnote{Beyond the quintic, new routines will be needed, as we will briefly discuss below.}. For concreteness, we use the homogeneous sampling method introduced in Section \ref{homsamp}, which gives a sample of points that is randomly distributed with respect to the Fubini--Study measure 
\begin{equation}
    d\, {\rm Vol}_{FS} = J_{FS} \wedge J_{FS} \wedge J_{FS} \; .
\end{equation}  
Second, in order to compare performance of different methods, we use standard error measures known as the $\sigma$ and Ricci measures. The $\sigma$ measure, first introduced by Donaldson, is obtained by integrating the Monge-Amp\`ere equation over the CY 
\begin{equation} \label{eq:sigma}
\sigma = \frac{1}{{\rm Vol}_{\Omega}} \int_X \left| 1- \kappa\; \frac{\Omega \wedge \overline{\Omega}}{(J)^n}\right|\; ,
\end{equation}
where  ${\rm Vol}_{\Omega} = \int \text{d} \vCY = \int \Omega \wedge \overline{\Omega}$ is the volume of the manifold. The Ricci measure is obtained by integrating the Ricci scalar
\begin{equation} \label{eq:ricci}
\mathcal{R} = \frac{{\rm Vol}_{CY}^{1/3}}{{\rm Vol}_{\Omega}} \int_X |R| \; .
\end{equation}
 These measures are closely related to the Monge-Amp\`ere and Ricci functionals introduced in \eqref{eq:MAfunc} and \eqref{Rfuncß}, and will be evaluated numerically as weighted sums, as discussed in Section \ref{homsamp}. 
 In general, the integral of a function $f$ is approximated by\footnote{The statistical error with computing integrals as such weighted sum is roughly $1/\sqrt{N}$, where $N$ is the number of points in the sample, as long as the weights/masses are order 1 numbers \cite{Douglas:2006rr} (this fails in highly curved regions, see \cite{Ahmed:2023cnw} for a discussion).  } 
	\begin{align}
		\label{eq:MCint}
		\int_X \text{d} \vCY f  = \int_X \frac{\text{d} \vCY}{\text{d}A} \text{d}A\; f  = \frac{1}{N} \sum_i m_i f |_{p_i} \quad \text{ with }\quad   m_i = \frac{\text{d} \vCY}{\text{d}A} |_{p_i}\;,
	\end{align}
where the weights/masses $m_i$ is the quotient between the CY measure and the FS measure. These weights should be computed during sampling, so that they can be used whenever an integration is to be computed.

Our main task then, is to devise a machine learning model that predicts the CY metric. This entails specifying what the model should predict:  should this be the metric $g$, the K\"ahler potential $K$, or the Hermitian matrix $H$ that defines the algebraic CY metrics of Donaldson's algorithm? All of these choices have been implemented in the literature, and their respective advantages will be discussed to some extent below. Given our choice of prediction, we then need to encode any remaining mathematical constraints as loss functions that guide the training of our network. Finally, we should determine the architecture of  our network (the type of layers, activation functions, etc.), and which training algorithm we should use. 

To keep this discussion within the time limit of these lectures we will now focus on the  implementation of the {\texttt{cymetric}} package \cite{Larfors:2021pbb,Larfors:2022nep}, which is constructed to directly predict the Ricci flat metric $g$. As such, this package includes and extends some of the implementations presented in \cite{Anderson:2020hux}. After this discussion, we will comment on alternative neural net designs, that are either provided by other packages, or that could be developed in the future.

\subsubsection{Approximate CY metrics from the  \texttt{cymetric} package}

Before describing the implementation, let us take a quick peak under the hood of the \texttt{cymetric} package. This will make our discussion more grounded, and hopefully give a feeling for how available ML libraries are used in this type of research.

The \texttt{cymetric} package is designed for CY metric prediction on generic CY manifolds, at any specified point in their moduli space (e.g. there is no need to restrict to points where the manifold is invariant under large symmetry groups). It decomposes in  point sampling routines and machine learning models.\footnote{The package is written in Python and Mathematica~\cite{Mathematica}, and we will predominantly refer to the Python version.} The point generators use functionality of \texttt{NumPy}~\cite{harris2020array} and \texttt{SciPy}~\cite{mckinney-proc-scipy-2010}, combined with routines from SageMath~\cite{sagemath} and Mathematica. The machine learning models  are implemented and optimized with TensorFlow~\cite{tensorflow2015-whitepaper} and Keras \cite{chollet2015keras}. This part of the package is structured into python classes, that inherit functionality from TensorFlow superclasses, and various helper functions that are used to call, compile and train models. The models are designed to be general, but care has been taken to write an efficient code, so that  ML experiments on non-trivial manifolds can be run without the need for large computing resources.\footnote{For example, the package uses the so-called function decorators of TensorFlow in order to speed up computations.}

\paragraph{Setting up the ML models}
Let us assume that we have generated a sample of points on the Fermat quintic. This provides the input to the neural network that we construct to predict the CY metric. In \texttt{cymetric}, the coordinates of each sampled point is saved as a tuple of twice the homogeneous dimension of the ambient space, i.e. $2*5$ for $\mathbb{P}^4$.\footnote{This tuple thus contains a redundant description of the three local complex coordinates of the CY; one reason for this design is that it allows to easily identify the coordinate patch the point belongs to; this information helps in constraining how the metric transforms between coordinate patches.} This takes into account that neural networks operate better with real numbers, rather than complex; an input layer with 10 nodes, can then read in the real and imaginary part of the coordinates. 

The output layer of the network should give the metric evaluated at the point specified by the coordinates we fed in to the input layer. This is naively a $6\times6$-dimensional matrix; however the Hermitian structure allows us to cut down the number of components to only predict to the 9 entries of a $3\times3$-dimensional, positive definite, Hermitian matrix (cf.~eq.~\eqref{eq:hmetric}). Thus, the output layer of our neural net should be 9-dimensional. 

In fact, the \texttt{cymetric} package allows the user to choose whether the NN should predict the full CY metric, as described in the last paragraph, or a correction to the FS metric  that is already known. The latter is likely more efficient, as a randomly initiated neural net may very well predict something very far from the CY metric (e.g. a negative definite matrix). The package provides a number of metric Ans\"atze of the type $g = g_{FS}+ ...$, as presented in table \ref{tab:ansatz}. The last of these, the $\phi$-model, differs in that it predicts the correction to the FS K\"ahler potential, not the FS metric; as such the output layer for a neural net with this Ansatz should be one-dimensional.
\begin{table}[t]
\centering
\begin{tabular}{@{}|c|c|@{}}
	\hline
	Name& Ansatz \\ \hline \hline
	Free & $g = g_{\text{NN}}$ \\
	Additive & $g = g_{\text{FS}} + g_{\text{NN}}$ \\
	Multiplicative, element-wise & $g = g_{\text{FS}} + g_{\text{FS}} \odot g_{\text{NN}}$ \\
	Multiplicative, matrix & $g = g_{\text{FS}} + g_{\text{FS}} \cdot g_{\text{NN}}$ \\
	$\phi$-model & $g = g_{\text{FS}} + \partial \bar{\partial} \phi$ \\
	\hline
\end{tabular}
\caption{Different ways to encode the Ricci-flat metric.}
\label{tab:ansatz}
\end{table}

Of course, specifying the input and output layer does not define a trainable neural network. We must define hidden layers, with parameters that can be adjusted during training. Observations in the literature show that simple,  shallow (2-3 hidden layers), fully connected, feed-forward networks trained by stochastic gradient descent work remarkably well. Setting up the neural net in \texttt{cymetric} is very simple, and uses the underlying TensorFlow/Keras functionalities. For example, by writing
\begin{python}
nn = tf.keras.Sequential()
	nn.add(tfk.Input(shape=(nIn)))
	for i in range(2):
    		nn.add(tfk.layers.Dense(nHidden, activation=act))
	nn.add(tfk.layers.Dense(nOut, use_bias=False))
\end{python}
we set up a dense feed forward neural net that consists of an input layer with \texttt{nIn} nodes, 2 hidden layers with \texttt{nHidden} nodes and non-linear activation function given by \text{act}, and an output layer with \texttt{nOut} nodes.  (see the \texttt{cymetric} package's GitHub page for the full example notebook). Thus, once the user has specified  hyperparameters such as the width, depth, and activation function, the ML libraries will do the rest of the initialization. This makes experimenting with different choices for these hyperparameters straightforward and the reader is encouraged to do so using the example notebooks.\footnote{When it comes to activation functions, it has been observed that GELU give better results than ReLU (possibly related to the problem of dying nodes of the latter), but there is room for experimentation here too.}. As already mentioned, simple NN designs give good results, but any design is possible; thus there is room to experiment by adding other types of layers when specifying the neural network.

Since the CY metric is unknown, the network is trained with the objective to minimize custom loss functions, rather than standard loss functions that encode the difference between the prediction and a known target. This is accomplished using backpropagation and stochastic gradient descent, 
just as described in Fabian Ruehles's first lecture of this school. In short, this training method computes gradients of the loss functions with respect to the network parameters, using the chain rule to propagate backwards from the last to the first layer of the network. The NN parameters are then updated in the direction of steepest gradient descent. This method has elaborations to avoid getting stuck in local minima, e.g. the parameter updates are done using smaller subsets (batches) of the input data (i.e. points on the CY), rather than the full data set. All of this functionality is provided by TensorFlow and Keras. In particular, derivatives of loss functions are efficiently computed using so-called gradient tapes. The ML libraries also provides several different optimizers that allow the user to specify elaborations of the backpropagation algorithm. 

 Once the neural net is trained, we can call it with the coordinates of any point of the CY, to get the approximation of Ricci flat metric evaluated at this point. Thus, the trained network is the (numerical approximation of the) CY metric. But how do we set up the loss function for this training? This is the topic of the next paragraph.

\paragraph{Training with custom loss functions}
Once we have set up the network, we need to build an ML model that we can train. To do so, we pick the metric ansatz of our choice from Table \ref{tab:ansatz}. At the initialization\footnote{By default, the parameters of the network are initialized to random numbers drawn from a Gaussian distribution (default is ${\cal N}(0,1)$ for the free network and ${\cal N}(0,0.01)$ for the other metric ansatze).} of the model, we must also specify which of the custom loss functions the training should be governed by.  

As mentioned above, the custom loss functions encode all the mathematical constraints that the CY metric should satisfy. Here, we have the  the Monge-Amp\`ere equation and the Ricci-flatness constraint. We also need to ensure that the predicted metric is K\"ahler, $d J =0$, and stays in the K\"ahler class of $J_{FS}$.  For some metric Ans\"atze, like the free and additive ones, it is also necessary to enforce more basic constraints such as a globally defined metric, which transforms correctly on patch overlaps in the atlas that covers the manifold.

In the \texttt{cymetric} package, these constraints are encoded in five custom loss functions $\mathcal{L}_i$, that combine to  
	\begin{align}
		\label{eq:loss}
		\cL &= \alpha_1 \cL_{\text{MA}} + \alpha_2 \cL_{\text{dJ}} + \alpha_3 \cL_{\text{transition}} + \alpha_4 \cL_{\text{Ricci}} + \alpha_5 \cL_{\text{Kclass}} \; .
	\end{align}
Here $\alpha_i$ are tuneable hyperparameters of the model (default value $\alpha_i = 1.0$) that are set by the user when calling the model. The minimization of these loss functions enforce that the predicted metric satisfies the MA equation, and is Ricci flat, K\"ahler, globally defined, and in the prescribed K\"ahler class. The first two loss function are, by the CY theorem, encoding the same constraint, so it should be enough to enforce only one of these during training.

In more detail, the
first four loss contributions are given by
\begin{itemize}
\item {Monge-Amp\`ere (MA) loss} $\cL_{\text{MA}}$\\
\begin{equation} \label{eq:mariloss}
\cL_{\text{MA}} = \left|\left| 1 - \frac{1}{\kappa} \frac{\det \gpred}{\Omega \wedge \bar\Omega}\right|\right|_n  \;  ,
\end{equation} 
\item {Ricci loss} $\cL_{\text{Ricci}}$\footnote{This is strictly only a good loss function for K\"ahler metrics, where the Ricci tensor simplifies to
$R_{i \bar{j}} =  \partial_{i} \bar{\partial}_{\bar{j}} \log \det (g)$.}\\
\begin{equation}
\cL_{\text{Ricci}} = ||R||_n = \left|\left|\partial \bar\partial \ln{\det\gpred}\right|\right|_n \;  ,
\end{equation}
\item {K\"ahler loss} $\cL_{\text{dJ}}$\\
\begin{equation} \label{eq:kloss}
\cL_{\text{dJ}} = \sum_{ijk}  \left|\left|\Re{c_{ijk}}\right|\right|_n +  \left|\left|\Im{c_{ijk}}\right|\right|_n \;  ,
\end{equation}
where $c_{ijk} = g_{i\bar{j},k} - g_{k\bar{j},i}$ and $g_{i\bar{j},k} = \partial_k g_{i\bar{j}}$. 
\item {Transition loss} $\cL_{\text{transition}}$\\
\begin{equation}
\cL_{\text{transition}} = \frac{1}{d} \sum_{\mathcal{U},\mathcal{V}} \left|\left|\gpred^{\mathcal{V}} - T_{\mathcal{U}\mathcal{V}} \cdot \gpred^{\mathcal{U}} \cdot \left(T_{\mathcal{U}\mathcal{V}}\right)^\dagger \right|\right|_n \; , \; 
\end{equation}
where $d$ is the number of patch transitions and the transition matrices $(T_{\mathcal{U}\mathcal{V}})_{\mu}^{\nu} = \partial v^\nu/ \partial u^\mu$  are the Jacobians of the coordinate transform between the coordinates $u$ in $\mathcal{U}$ and $v$ in $\mathcal{V}$. 
\end{itemize}
These losses are computed over mini-batches containing $N_B$ points, and as indicated by the subscript $n$, the loss functions can be computed with any $L_n$ norm\footnote{Default in the package is $n = 1$ for all losses but  $\cL_{\text{dJ}}$, which has $n=2$. One should choose the $L_n$ norm wisely, as it affects the training of the network. With $n=1$, all points in the batch are treated equally, and the network strives to minimize the average loss. With higher $n$, the network prioritizes outliers, i.e. points with the largest loss function contributions.}:
\[
||x||_n = \left( \sum_{i=1}^{N_B} x_i^n \right)^{1/n}
\]
For all these loss functions, the derivatives with respect to the input coordinates are computed with TensorFlow's automatic differentiation, but with the twist that gradient tapes are applied to the input, not network parameters. There is a computational cost associated to this, in particular for the Ricci loss which requires evaluating two derivatives. Therefore, this loss function, which is in principle encoding the same constraint as the MA loss on CY geometries, is disabled by default in the package.

There is a final component in the loss, which enforces that the predicted metric stays in the correct K\"ahler class. Since this loss function is of little interest for experiments on the quintic, or any other CY manifolds with $h^{(1,1)}=1$, we defer the description of this loss function to the next section.

With these loss functions defined, we can initialize and train the ML model using the full functionality of TensorFlow. This is simply done by first initializing the chosen model with the neural net, informaton about CY geometry, and loss function hyperparameters as inputs. E.g.
\begin{python}
    fmodel = AddFSModel(nn, BASIS, alpha=[1., 1., 1., 1., 1.])
\end{python}
would initialize an ML model corresponding to the additive metric ansatz, with default values of the loss hyperparameters $\alpha$, which has access to the information of the CY contained in \texttt{BASIS}.\footnote{This data container is created by the point generators of \texttt{cymetric}, and is described in detail in the package's documentation. It holds information such as the constant $\kappa$, the defining polynomial for the CY, etc.}

Second, we compile the model, and then train it using the custom loss function on the sampled points. This is done using the \texttt{compile} and \texttt{fit} commands of TensorFlow/Keras. To simplify the computation of the various components of the loss, we call these functions using the \texttt{train-model} helper function of the \texttt{cymetric} package, i.e.
\begin{python} 
    fmodel, training_history = train_model(fmodel, data, optimizer=opt, epochs=N, batch_sizes=[64, 50000], verbose=1, custom_metrics=cmetrics, callbacks=cb_list)
\end{python}
This command starts a training loop that will run over N epochs, where each epoch consists of a first training loop over all mini batches (containing 64 points) and a second training loop over the larger batches (containing $50,000$ points). This two-step training will, as alluded to above,  first update the network parameters to minimize the MA, K\"ahler, and transition losses.\footnote{Recall that the Ricci scalar is disabled by default.}  For the quintic, the second training step (over the large batches) is unnecessary as its objective, preservation of the  K\"ahler class, is automatic.\footnote{To avoid this unnecessary step, the \texttt{compile} and \texttt{fit} commands may  be used directly. Alternatively, the user may modify the \texttt{train-model} function by removing the second training step. For the experiments done below, the second training step only adds a second per epoch, in comparison to the 30s needed for the first step. Thus the time saved is not significant, but for longer training on more complicated Quintics, it may be. } We will describe this loss function in  more detail in the next section.

\begin{figure}[htb]
    \centering
    \includegraphics[width=
\textwidth]{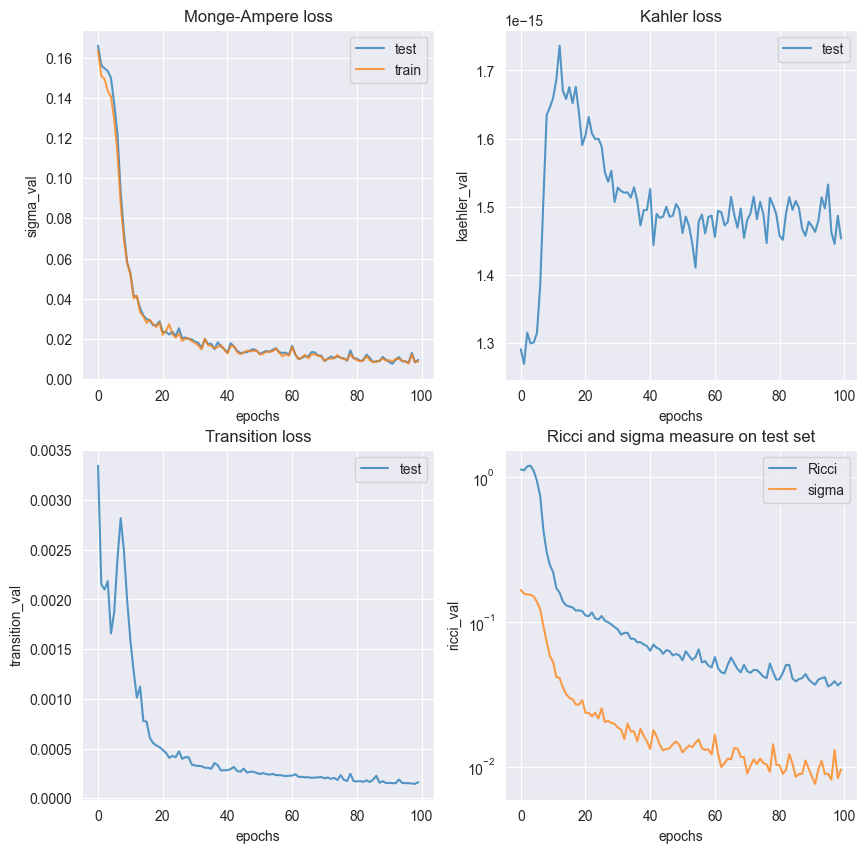}
    \caption{A simple experiment using the $\phi$ network on the Fermat quintic. Training clearly reduces the MA and transition loss, and keeps the Kahler loss negligible. The error measures, computed on points not seen during training, also decrease.}
    \label{fig:Fermatquintic}
\end{figure}

\begin{figure}[htb]
    \centering
    \includegraphics[width=
\textwidth]{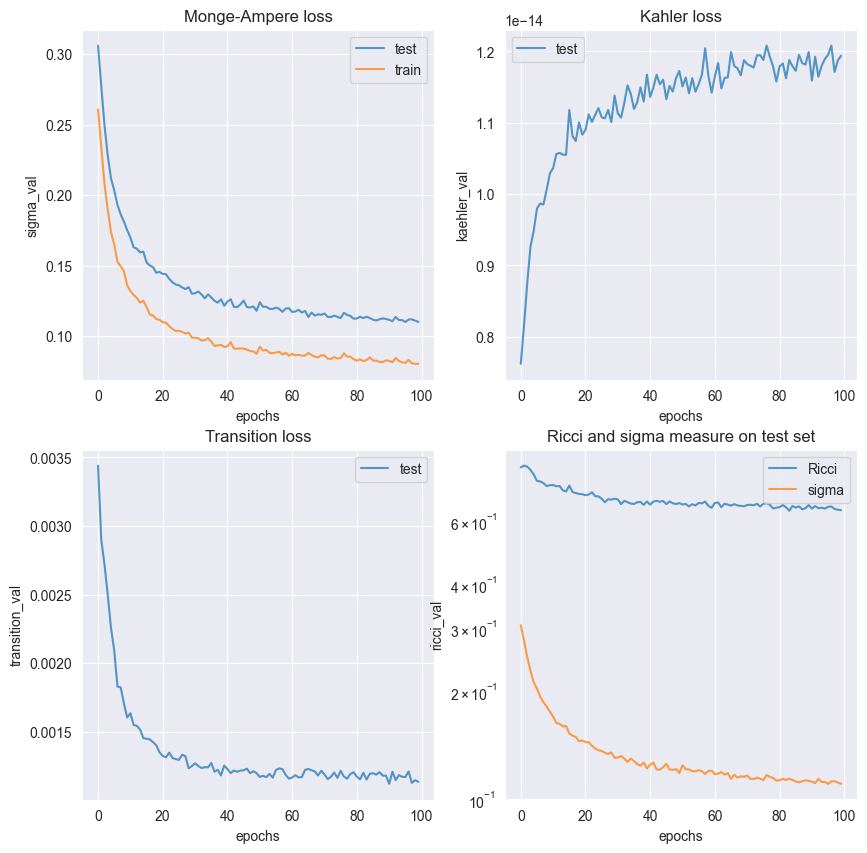}
    \caption{Training the $\phi$ network on a quintic at a non-symmetric point in moduli space. Training takes slightly longer than on the Fermat quintic (notice the non-vanishing slope in the MA loss), and the results don't generalise as well to the test set. However, qualitatively, the plots are very similar; clearly learning is taking place also on this manifold.}
    \label{fig:nonsymquintic}
\end{figure}

As evident from the code snippet above, when we train the model, the compile and train commands must also specify callbacks and custom metrics, which allows us to track the training dynamics. The package's custom metrics consist of the loss functions (including the Ricci loss), so that we can check that these decrease during the training process. The callbacks are measures of the accuracy of the metric prediction that we will evaluate on test data. These include the $\sigma$ and Ricci measures in eqs. \eqref{eq:sigma} and \eqref{eq:ricci}, which are  benchmark measures~\cite{Douglas:2006hz,Anderson:2010ke} used to compare the accuracy of any numerical approximation of the CY metric, including the algebraic metrics that result from Donaldson's algorithm or Headrick and Nassar's functional methods. In addition to these standard measures, there are also callbacks that check the constancy of the volume, and that the metric prediction is K\"ahler and globally defined. These callbacks are needed to monitor the machine learning implementation of the \texttt{cymetric} package, where these are soft constraints, that are imposed through the loss function.  

To conclude, we have described what we need to do to machine learn Ricci-flat CY metrics. To summarize, we must
\begin{enumerate}
    \item Generate points
    \item Create NN
    \item Choose ML model (i.e.~metric Ansatz)
    \item Train model
    \item Check accuracy
\end{enumerate} 
Let's gauge the computational resources such experiments takes. On a standard laptop\footnote{To be more precise, on a 2016 MacBookPro that has seen better days.}, generating some 100,000 points takes from a few minutes to half an hour for CICY manifolds with moderate number of K\"ahler moduli, and a bit longer for CYs in toric ambient spaces. Training the model for 50 epochs takes half an hour on simple Quintics, and in the order of hours for more elaborate CYs; as one would expect, the simpler (more symmetric) the geometry, the faster the training. Using a machine with a GPU leads to faster execution.

\paragraph{Simple experiments}
We are now ready to perform experiments using the \texttt{cymetric} package. Let's start simple, and run one of the ML models on the Fermat quintic. We will pick the $\phi$ model with default loss function weights. After generating $100,000$  points on the manifold (which we split 9:1 into a training and validation set, where the latter is hidden during training), we set up a fully connected NN with 3 hidden layers of width 64. We select GELU activation functions, the Adam optimizer, and train for 100 epochs (about 1 hour). The results of the training are given in Figure  \ref{fig:Fermatquintic}. We see that the MA loss decreases, during training, by a factor 20. This is true both for the training set and the validation set. We see that the Kähler loss is vanishing throughout the training and that the transition loss functions decreases by a factor of 10 (both evaluated on the validation set). Finally we plot how the Ricci  and sigma measures decrease during training, reaching a level of 0.05 and 0.01 respectively. These training results clearly shows that learning is taking place. 

Note that we have looked at a very simple network, with the most standard set up for the training. The models may be optimised in many different ways --- indeed, the non-zero slopes of the losses and error measures even indicate that simply training longer will improve the results.  The results are nevertheless proof that the ML methods work (and that work towards better precision is motivated, as we will discuss below).

What is more, these ML methods have a wide range of applicability. We can easily run the experiment that we just did, but for a quintic defined by a different  polynomial. As an example, we can construct a completely generic quintic. Following \cite{Braun:2008jp}, we   pick random, complex coefficients $a_i$, of norm $|a_i|\in (0,1)$  for the 126 homogeneous quintic monomials
\begin{equation}
\begin{split}
p = \sum a_i m(Z) = a_0 (Z^0)^5 + a_1 (Z^0)^4 Z^1 + ... a_{126} (Z^4)^5 \; ,
\label{eq:pnonsym}
\end{split}
\end{equation}
This breaks all discrete symmetries of the quintic, and thus using traditional numerical methods to compute the Ricci-flat metric would require extensive computational power.

We again generate  $100,000$  points on this non-symmetric quintic. Running the exact same experiment as for the Fermat quintic takes again of the order of hours. The results of the training are presented in Figure \ref{fig:nonsymquintic}. Qualitatively, these look very similar to the Fermat results: the  loss functions and error measures decrease during training. Quantitatively, the values for the MA and transition losses are larger than in the Fermat case, and the loss decrease is not as big (factor of 5 rather than 20 for MA and transition loss; the Kahler loss is still vanishing). The gain on the error measures is a factor of 2-3.  

 While it is hard to draw  conclusions from these experiments without more systematic study,\footnote{Recall the large amount of randomness inherent in these experiments; we sample random points, we train with stochastic gradient descent, etc., and so fluctuations in performance are expected. It is good practice to run a number of different experiments, and draw conclusions based on average performance.  } one may observe that the ``non-symmetric" model generalizes less well to the test  set. This difference in generalizability is likely attributed to  symmetry. Recall that the Fermat quintic is a highly symmetric manifold. Its defining polynomial is invariant under discrete $\mathbb{Z}_5$ rotations, permutations, and complex conjugation of the homogeneous coordiniates.\footnote{Taking the projective rescaling into account, the full symmetry group of the Fermat quintic is $\mathbb{Z}_5^4 \rtimes (S^5 \times \mathbb{Z}_2)$ \cite{Braun:2008jp}.} This symmetry must be learned to some extent by the network, and this will help it to generalise to new points on the manifold. On the other hand, the quintic defined by \eqref{eq:pnonsym} breaks these symmetries. As such, the network must approximate a more complicated metric, and will likely be more complicated and possibly more prone to fitting to non-generalisable features of the training data.  Interestingly, despite this difference in training results, the final value for the Ricci and sigma measures are comparable in the two  experiments. This indicates that the ML performance is similar on the two quintics.

The main point of performing this second experiment on the non-symmetric quintic is that we are no longer limited to working with highly symmetric CY manifolds. Indeed, the direct machine learning approach works also far beyond the regime accessible for algebraic metrics. With machine learning, we can look at any example from the CY databases, and this is very exciting for string theory (and mathematics).

\subsection{Machine learning experiments beyond the quintic}

Most CY manifolds in the CICY and Kreuzer--Skarke databases differ from the quintic in one of the following respects: they may have $h^{(1,1)} > 1$, be defined by more than one homogeneous polynomial, and/or be defined as a hypersurface in a toric ambient space, rather than a projective one. This calls for new point generators and a new training loop. Such point generating algorithms have been developed for CICY manifolds \cite{Braun:2007sn} and CY manifolds in toric ambient spaces \cite{Larfors:2021pbb,Larfors:2022nep}. They are direct generalisations of the homogeneous sampling technique described for the Quintic sin section \ref{homsamp}, and are included in the \texttt{cymetric} package's point generator routines.  
Describing how this works in detail would require another lecture, so instead we will refer the reader to \cite{Larfors:2022nep}, where this topic is discussed in detail. We will instead focus on the training loop modifications that are needed for more elaborate CY manifolds.

For manifolds with more than one K\"ahler class, there is a risk that training will push the predicted metric out of the initial K\"ahler class. This is problematic: the constant $\kappa$ is computed in this K\"ahler class, and the MA equation loses its meaning if the network jumps between classes. To ensure this does not happen, a new loss function must be added. In the \texttt{cymetric} package, the following loss is defined:
\begin{itemize}
\item {K\"ahler class loss} $\cL_{\text{Kclass}}$\\
\begin{equation} \label{eq:tvolkloss}
\cL_{\text{Kclass}} = \frac{1}{h^{1,1}(X)}\sum_{\alpha=1}^{h^{1,1}(X)}\left|\left|\mu_{t}({\cal O}_X(e_\alpha))-\int_X J_\text{pr}^2\wedge F_{FS,\alpha}\right|\right|_n \; .
\end{equation}
This function computes certain topological quantities, the slopes $\mu_t$, for a basis of line bundles ${\cal O}_X(e_\alpha)$, $\alpha=1,\ldots ,h^{1,1}(X)$. These line bundles are specified by integers $(e_\alpha)^\beta=\delta_\alpha^\beta$, and their slopes $\mu_{t}({\cal O}_X(e_\alpha))$ are topological quantities which depends on the choice of K\"ahler class through the K\"ahler moduli $t^\alpha$. They can be computed using the intersection numbers of the manifold.\footnote{The intersection numbers are topological invariants of the CY manifold; they are known for CICY and KS CY manifolds.} Being topological, these quantities should be the same for all metric choices. The slope may also be computed as the integral after the minus sign in \eqref{eq:tvolkloss}, where the $J_\text{pr}$ is the predicted K\"ahler form, and  $F_{FS,\alpha}=-i J_\alpha/(2\pi)$ are the line bundle field strengths computed using the Fubini-Study K\"ahler form.\footnote{For CY manifolds with more than one K\"ahler modulus, the Fubini-Study K\"ahler form is given by $J_{FS}= \sum t^\alpha J_\alpha$, where $t^\alpha$ are the K\"ahler moduli.}
This integral is evaluated numerically as a weighted sum over the sampled points. This is not sensible to evaluate on a mini-batch of around $100$ points. Instead, it should be evaluated on a large batch. 
\end{itemize}

With this addition to the loss function, we can learn Ricci flat metrics on CY manifolds with multiple K\"ahler moduli. Given a point sample, we set up a neural net for our preferred metric Ansatz as described above. Subsequently, we compile and train the model on the sampled points using the same \texttt{train-model} command. This starts a training loop that runs over N epochs, using the optimizer, custom metrics and callbacks we choose to specify. During training, each epoch consists of a first training loop over all mini batches (containing 64 points) and a second training loop over the larger batches (containing $50,000$ points). This two-step training will  first update the network parameters to minimize the MA, K\"ahler, and transition losses, and secondly, the Kclass loss.\footnote{Recall that the Ricci scalar is disabled by default.} The optimizer (with specified learning rate and momentum) is shared  between the two optimization steps, and gradients are accumulated in the training.%

The double training loop, with it's numerical slope computation, makes training a bit slower than using TensorFlow's standard command.  However, there is no way around this: we must ensure the K\"ahler class is preserved in order to solve the MA equation, and the Kclass loss of the \texttt{cymetric} package is, at the time of writing these notes, the only way to ascertain this will be true on CY manifolds with $h^{(1,1}>1$.

\subsection{Outlook and open problems}

In these lectures we have described numerical methods that can be used to approximate Ricci flat CY metrics. 
For the machine learning implementations, we have centered our discussion around one particular tool, the \texttt{cymetric} package. With this package, we can compute approximate CY metrics on any CY manifold in the CICY and KS databases, at any given point in their moduli space. This package encodes the metric in a neural network, which is trained on a random sample of points on the CY manifold. After training, the metric can be computed at any given point on the manifold by simply calling the trained ML model. 

As already mentioned there are two other open-source packages, \texttt{MLgeometry} \cite{Douglas:2020hpv} and \texttt{cyjax} \cite{Gerdes:2022nzr}, that also may be used to predict CY metrics. These packages are similar to \texttt{cymetric} in that they use fully connected neural networks, trained by stochastic gradient descent. They differ in some architectural choices for the neural nets, the ML libraries they rely on, and in how points are sampled and presented to the ML models. One important difference is that they encode more of the Calabi--Yau constraints in the design of the neural net. Rather than predicting the metric, these networks predict the K\"ahler potential $K$, or Donaldson's $H$ matrix. Second, global consistency is ensured by expanding $K$ in homogeneous polynomials of degree $k$, similar to what is done for the numerical methods using algebraic metrics. With this alternative design, there is no need to encode global consistency or K\"ahlerity  constraints as loss functions; in fact only the MA loss is used in the training of these networks.

The encoding of constraints in the network is advantageous, when it comes to the accuracy of the prediction, a topic discussed in e.g.~\cite{Douglas:2020hpv}. It has also made it possible to explore the complex structure moduli dependence of the CY metric approximation \cite{Anderson:2020hux, Gerdes:2022nzr}. However, encoding more information in the network leads to designs that are harder to generalise to new settings. Indeed, these models are, to date, tailor made for CY spaces with a single K\"ahler class. Thus, they  require some adaption to work on general CY manifolds. One disadvantage in such generalisations would then be the curse of dimensionality associated with the dimension of the polynomial basis, which will be the same as in the ``pre-ML" numerical methods.  An interesting middle ground, dubbed the spectral network,  was recently proposed in \cite{Berglund:2022gvm}. Here the general implementation of \texttt{cymetric} package is combined with an expansion of the metric in a basis of homogeneous polynomials. This is shown to improve the accuracy of metric predictions, in particular in nearly singular quintic  manifolds, in agreement with the findings of \cite{Douglas:2020hpv}. 

As we are concluding these lectures it is reasonable to ask how well these machine learning methods perform relative to their more conventional counterparts? Such comparisons have been made in the literature, usually  by comparing the ML methods to the Donaldson algorithm (see e.g. \cite{Anderson:2020hux,Douglas:2020hpv}). However, as was described in Section \ref{sec:perf&prob}, standard functional minimization already achieves essentially the optimal metric, which is closest to Ricci-flat for the given ansatz, up to errors that are negligible compared to working at finite $k$. Thus one might think that at a given $k$ machine learning methods can not do any better. This has been shown not to be correct for two reasons. Firstly, as already explained, ML methods can compute these metrics without using the symmetry assumptions that are required in the conventional approach. We can thus perform computations in a couple of hours, or even a few minutes in optimized settings, at points in moduli space that simply do not finish with conventional techniques. Secondly, some results which have been obtained in the machine learning contexts have been computed as a function of complex structure moduli \cite{Anderson:2020hux,Gerdes:2022nzr}. This too has never been achieved with more conventional approaches. Thirdly, with alternative network architectures, training of ML models can speed up significantly, see \cite{Halverson:2023ndu} for an exciting new proposal that, in very short time, provides highly accurate approximations for the Ricci flat metric on the training data from the Fermat quintic. However, the models of this reference do not generalise well; in the current form, they are not providing accurate predictions on test data. 

Hopefully,  these lectures have demonstrated that numerical and machine learning methods are useful in exploring CY geometries. Undoubtedly, there is also scope to improve these methods further. Indeed, it is important to stress that this is a field of research where there are many open questions. In particular, the field of ML and CY geometry is still in its infancy and breakthroughs regarding efficiency, accuracy, and scope are likely to come. One very interesting open question is how to approximate moduli-dependent CY metrics beyond the partial explorations of the quintic's complex structure moduli space done in Refs \cite{Anderson:2020hux, Gerdes:2022nzr}. Another interesting application is to use numerical CY metrics to identify regions of high curvature on CY manifolds, at different points in their moduli space, see \cite{Cui:2019uhy, Berglund:2022gvm} for recent discussions. 

Numerical methods also allow us to compute massive spectra of string compactifications, with relevance to e.g. the Swampland program, see \cite{Braun:2008jp} and \cite{Ashmore:2020ujw,Ashmore:2021qdf,Ahmed:2023cnw}. These numerical methods also allow us to compute masses and couplings related to vector bundles over CY manifolds, see \cite{Anderson:2010ke,Anderson:2011ed} and \cite{Ashmore:2023ajy}. In a different direction, numerical and machine learning methods have only very rarely been applied outside the CY regime. Other geometries are also of interest in string theory, such as   different types of $G$ structure geometries. As demonstrated in \cite{Anderson:2020hux}, one can make progress in some of these direction using the ML techniques described here. Indeed, with general ML models, one need only adapt the loss functions  to encode, instead of the CY constraints, whatever restrictions such metrics should satisfy. In \cite{Anderson:2020hux} this allowed the authors to machine learn an SU(3) structure on the quintic which was shown to exist in \cite{Larfors:2018nce}. 

Finally, returning to the CY case, it is fascinating that the simplest type of neural networks can make progress on such a non-trivial mathematical problem. Does this mean that Ricci flat CY metrics are, perhaps, simpler than we have thought? Perhaps, by analysing models trained on different CY geometries, we can  hope to detect universal properties of their  Ricci flat metrics? Perhaps, you will contribute to answering this question (hopefully in the positive) in the future?

\section*{Acknowledgement}
These notes are a contribution to the  LMS Research School: Machine Learning in Mathematics and Theoretical Physics 
(organised in Oxford in summer 2023 by Andrei Constantin and Yang-Hui He), to appear in the book ``ML Tutorials in Pure Mathematics and Theoretical Physics'' (World Scientific). The authors would like to thank the organisers and participating students of this school. Lara Anderson and James Gray are supported by NSF grant PHY-2310588. Magdalena Larfors was in part supported by Vetenskapsrådet, grant 2020-03230, and Uppsala University's Ai4Research center. Her work was partially supported the Wallenberg AI, Autonomous Systems and Software Program (WASP) funded by the Knut and Alice Wallenberg Foundation

\bibliographystyle{alpha}
\bibliography{refs}

\end{document}